\documentclass[lineno]{jfm}

\usepackage{graphicx}
\usepackage{newtxtext}
\usepackage{newtxmath}
\usepackage{natbib}
\usepackage{comment}
\usepackage{amsfonts,mathtools,amssymb,stmaryrd,dsfont}
\usepackage{hyperref}
\hypersetup{
    colorlinks = true,
    urlcolor   = blue,
    citecolor  = black,
}

\newcommand{\RomanNumeralCaps}[1]
\linenumbers

\newcommand{\so}{\color{black}}
\newcommand{\os}{\color{black}}

\newcommand{\si}{\color{black}}
\newcommand{\is}{\color{black}}

\newcommand{\sii}{\color{black}}

\title{\so Meltwater  transport \os and mixing layer growth near the ice\textendash ocean interface}
\author{Sofía Allende\aff{1}
  \corresp{\email{sofia.allende@univ-amu.fr}},
Louis-Alexandre Couston\aff{2},
Simon Thalabard\aff{3},
 \and Benjamin Favier \aff{1}}

\affiliation{\aff{1}Aix Marseille  Univ, CNRS, Centrale Med, IRPHE, Marseille, France.
\aff{2}Université Claude Bernard Lyon 1, ENS de Lyon, CNRS, LPENSL, UMR5672, 
Lyon, % cedex 07, 
France.
\aff{3}Université Côte d'Azur, CNRS, Institut de Physique de Nice, 
Nice, France.}

\begin{document}
\maketitle

\begin{abstract}
Ice melting into saline water plays a fundamental role in the dynamics near the  ice-ocean interface in  polar oceans. \so The physics of meltwater transport  involves \os a non-trivial interplay between thermodynamics at the  interface,  hydrodynamic transport in the bulk and the properties of the ambient ocean.  \so The key control parameters are the density ratio $R_\rho$, which is proportional to the ambient salinity and measures the balance between the temperature and salinity effects on density, together with
 the Lewis number $Le = \kappa_T/\kappa_S$,  which compares  thermal and solutal diffusivities. 
\so In quiescent horizontal configurations, increasing the  salinity  is known to  slow down melting, 
with the melt rate transitioning from subdiffusive to diffusive as $R_\rho$ increases. \os
Here, we  assess the role of turbulence in this transition, using highly-resolved numerical simulations of the two-dimensional Boussinesq equations  with a  slowly melting upper boundary. We analyse the non-stationary growth of the  \so thermal and solutal \os
mixing layers, varying the Lewis number and  the density ratio. 
 While meltwater transport is continuously  driven by convection within the bulk, we  identify a transition from convection to diffusion close to the interface. This transition is reflected by the formation of an interfacial boundary layer that  regulates the flux of meltwater pouring into the turbulent bulk.
\so The boundary layer has little effect on thermal transport, but strongly suppresses  solutal transport for $R_\rho \gtrsim 10$, only allowing a fraction $\propto R_{\rho}^{-1}$ of the input flux to reach the bulk.   \os 
Using mixing-layer diagnostics based on  \so solutal-concentration thresholds\os, we observe that the turbulent layer grows  super-diffusively  $\propto t^{1.33}$, while the interfacial  boundary layer expands diffusively $\propto t^{0.5}$ but with a non-universal prefactor depending on Lewis and density ratio.
\sii The superdiffusive growth of the mixing layer challenges the commonly assumed picture of a fully diffusive regime at high salinity. 
%\so In particular, the mixing layer grows super-diffusively in all regimes, thereby challenging the commonly assumed picture of a fully diffusive regime at high salinity.
Overall,  our \os results indicate that  double-diffusive effects are here confined  to the interface, and highlight  potential limitations of diagnostics based on fixed concentration thresholds in oceanographic applications.
\end{abstract}

%\begin{keywords}
%Fresh-water melting, Mixing length, Turbulent fluxes. %Double diffusive convection
%\end{keywords}

%{\bf MSC Codes }  {\it(Optional)} Please enter your MSC Codes here

\section{Introduction}
\label{sec:intro}
The melting of ice into salty water is a fundamental process that occurs in various natural environments,  controlling the fate  of polar ice shelves, sea ice, and glacial meltwater plumes~\citep{hewitt2020subglacial, du2024physics, rosevear2025does}. 
 Understanding its process is  critical for predicting sea level rise and climate feedback mechanisms, as melting affects both local stratification and the transport of freshwater into the ocean \citep{jenkins1999impact}.
 Recent  studies have emphasised the sensitivity of melting rates to thermodynamics at the interface~\citep{keitzl2016impact}, examined fluid instabilities caused by basal diffusion 
\citep{berhanu2021solutal,cohen2020buoyancy}, explored the role of turbulent mixing in influencing the properties of near-ice water \citep{couston2024turbulent}, and investigated the asymmetries between melting and freezing within convective systems \citep{yang2025asymmetric}. 

The recent numerical studies by~\citet{xue2024flow} and~\citet{guo2025effects}, employing phase-field methods, 
suggest  that in the absence of external flow, the interface dynamics is controlled by the density ratio $R_\rho$
\so a non-dimensional number later defined in Eq. ~\eqref{eq:numbers}, which is proportional to ambient ocean salinity and reflects the balance between the temperature and salinity effects on density. \os 
 Over relatively fresh ambient water (small $R_\rho$), mixing is  dominated by thermal convection, and ice melts \sii relatively rapidly in time, 
%\so slowly in time
 following a sub-diffusive scaling $\propto t^{-r}$ with the exponent $r \lesssim 0.2$.  \so Over relatively salty ambient water (large $R_\rho$), the system instead develops a strongly stratified, diffusion-dominated regime, with   the melt rate exhibiting a diffusive scaling $\propto t^{-0.5}$, as also reported by~\citet{middleton2021numerical}\os.

Here, we revisit  the turbulence generated by a block of freshwater ice immersed in  initially quiescent, warm,  salty  ocean water. 
\so  In \citet{xue2024flow} and \citet{guo2025effects},  the physics of the melting interface is treated in considerable detail, making nonlinear effects away from the interface difficult to isolate.  Our setting explores the opposite limit: a simplified modeling of the interface, allowing nonlinear effects away from it to be more clearly assessed. \os
Our study focuses on the  early-time growth of the mixing layer (\emph{i.e.}, while unaffected by the bottom boundary) and its connection to the melt rate at the ice\textendash ocean interface. 
We use high-resolution numerical simulations  \so of the Boussinesq system under a linearised equation of state, \os inside a  vertically bounded and horizontally periodic two-dimensional domain\so---supplemented by a slowly melting condition at the top boundary. \os We investigate how the Lewis number (the ratio of thermal to saline diffusivity) and the density ratio impact the dynamics. \so 

In our configuration, a convective mixing layer develops,  which grows super-diffusively $\propto t^{1.33}$ and self-similarly in time. The growth rate is largely universal, as it  does not depend \sii on double diffusive effects, which are absent  in the bulk, nor on the melt rate at the top boundary. \os
\so
Over sufficiently saline environments, we recover the formation of a meltwater boundary layer, which produces a diffusive regime  of melting. The interfacial layer does not suppress  the turbulence inside the convective mixing layer, nor does it significantly modify  its statistics. However, it regulates the amount of meltwater  poured into the bulk: we find that the  ratio between  the convective and diffusive contributions to the solutal fluxes decreases at a rate  $\propto (LeR_\rho)^{-1}$, depending both on  the Lewis and the density ratio.
Because of its relative simplicity, our setting  allows for some theoretical understanding of the meltwater transport, from  the diffusive to the convective stages.
\os

The paper is organized as follows: \S \ref{sec:method} describes the physical and numerical setup, 
 \S\ref{sec:results} presents the main results, contrasting interface and bulk dynamics and discussing mixing length definitions. Finally in  \S\ref{sec:conc}, we share some concluding remarks. \so Additional technical details related to numerical setups and theoretical arguments are described in Appendix. \os

%######################################################################
%######################################################################
%######################################################################
% SECTION: NUMERICAL SET-UP
\section{Numerical set-up}
\label{sec:method}
\begin{figure}
    \centering
%    \hspace{0.02\textwidth}
   \begin{minipage}[t]{0.48\textwidth}
        \centering
        \includegraphics[width=\linewidth]{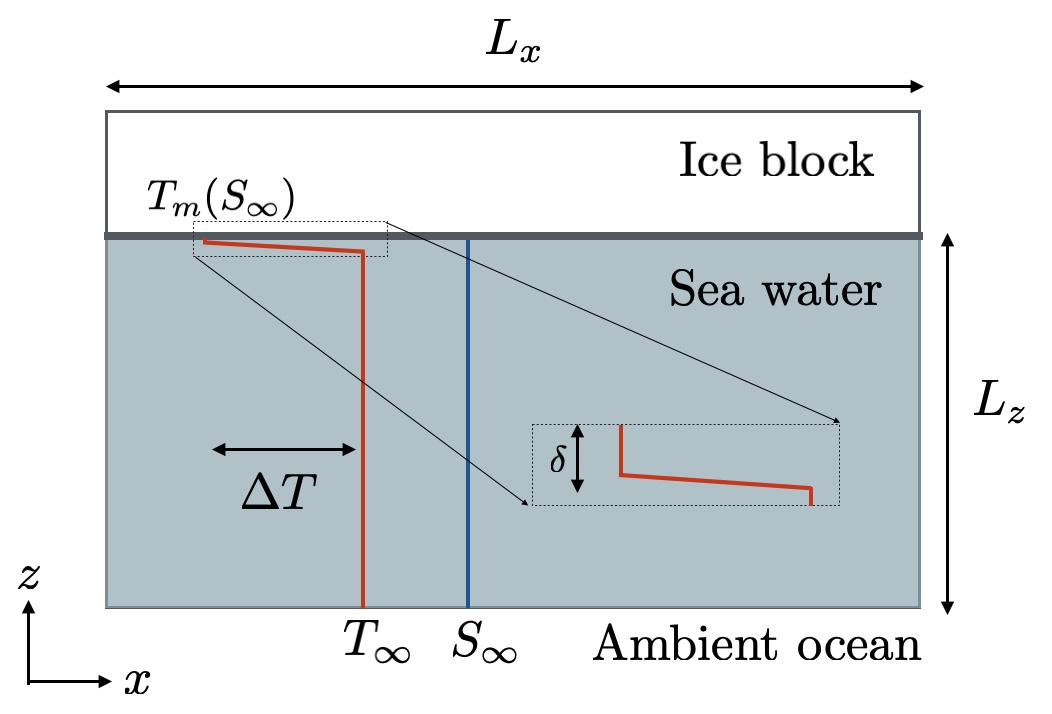}
        \caption{Initial configuration}
        \label{fig:scheme}
    \end{minipage}%
   \begin{minipage}[t]{0.48\textwidth}
	\scriptsize
    	\vspace{-4.5cm}
        \centering
        \begin{tabular}{lll}
           \hline
            Heat capacity of seawater & $C_p$ &  $3974 ~J K^{-1} kg^{-1}$\\
            Latent heat  &$L_f$ & $3.35 \times 10^{5} J kg^{-1}$ \\
            Freezing slope & $\lambda_1$ & $-5.73 \times 10^{-2}~ K (g/kg)^{-1}$ \\
            Thermal expansion & $\beta_T$ & $3.87 \times 10^{-5}~ K^{-1}$  \\
            Haline contraction  & $\beta_S$ & $7.86 \times 10^{-4}~ (g/kg)^{-1}$ \\
            Temperature jump & $\Delta T$ &  $2~K$\\
          %  \so Salinity jump & $\Delta S$ &  $0.098$ \os \\
            Freezing point of pure water & $T_{{f,0}}$ & $273.15~K$  \\
	   Kinematic viscosity & $\nu$ & $1.8 \times 10^{-6}~m^2s^{-1}$\\
	  Thermal diffusivity & $\kappa_T$ & $1.8 \times 10^{-6}~m^2s^{-1}$\\
	  Gravitational acceleration & $g$ & $9.8~ms^{-2}$\\
            Depth  & $L_z$ & $6.28~m$ \\
            Horizontal extent  & $L_x$ & $6.28~m$  \\
            \hline
        \end{tabular}
        \captionof{table}{Physical parameters }
        \label{tab:therm_param}
    \end{minipage}
\end{figure}

We consider the  flow illustrated in Fig.~\ref{fig:scheme}, which is initially quiescent. The domain is two-dimensional (2D), with
horizontal extent $L_x$  and depth $L_z= L_x$. The initial salinity $S$ is uniform and equal to the far-field value $S_\infty$, while the temperature $T$ has  a step-profile characterized by a temperature jump $\Delta T = T_\infty - T_m(S_\infty)$. This represents the difference between the far-field temperature $T_\infty$ and the melting temperature $T_m(S_\infty)= T_{f,0} + \lambda_1 S_\infty$. 

We later work with non-dimensional variables for temperature and salinity, defined  as
\begin{equation}
\begin{split}
 \theta&= \dfrac{T-T_\infty}{\Delta T}  \quad \text{and} \quad \sigma =  \dfrac{S-S_\infty}{\Delta S}\\
\text{where }\quad & \Delta T:= T_\infty-T_m(S_\infty), \quad \Delta S :=\dfrac{\beta_T}{\beta_S} \Delta T. 
\end{split}
\label{eq:rhosigma}
\end{equation}
 Here, $\Delta S$ is the salinity variation required to compensate the density change caused by $\Delta T$, according to the  linearised equation of state  
\begin{equation}
\rho = \rho_0 \left[1- \beta_T T + \beta_S S \right],
\label{eq:EoS}
\end{equation}
 which we adopt here for simplicity.
We model the fluid dynamics with \is the two-dimensional incompressible Navier-Stokes equations under the Boussinesq approximation. 

Defining  the horizontal length $L_x$, the velocity   $U_0 :=\sqrt{2 g \beta_T \Delta T  L_x }$, and the time  $t_0= L_x/U_0$,  as characteristic scales, the \so equations of motion \os%dynamics 
take the dimensionless form
\begin{equation}
\begin{split}
\partial_t \mathbf{u} + (\mathbf{u} \cdot \nabla) \mathbf{u} +\nabla p & =  Re^{-1}  \Delta \mathbf{u} + \mathbf{e_z} \frac{\theta-\sigma}{2}, \quad \quad \nabla \cdot \mathbf{u} = 0, \\
\partial _t \theta + (\mathbf{u} \cdot \nabla) \theta &= Re^{-1} \, Pr^{-1} \Delta \theta, \\
\partial _t \sigma + (\mathbf{u} \cdot \nabla) \sigma &= Re^{-1} \, Pr^{-1}  \, Le^{-1}   \Delta \sigma,
\end{split}
\label{eq:ns-boussinesq}
\end{equation}
with 
 non-dimensional parameters 
\begin{equation}
Re = \dfrac{U_0 L_x}{\nu}\text{(Reynolds)},\quad Pr=\dfrac{\nu}{\kappa_T}\text{ (Prandtl)} \quad \text{and} \quad Le = \dfrac{\kappa_T}{\kappa_S} \text{ (Lewis)}.
\end{equation}
The Prandtl  and the Lewis numbers represent  the ratio of viscosity to thermal diffusivity and  thermal to salt diffusivity, respectively. 

The boundary conditions represent a (no-slip) melting ice\textendash ocean interface at the top coupled to a far-field quiescent state at the bottom, similar to \sii \cite{couston2024turbulent}. 
At the top boundary, the temperature $\theta$ is set equal to the \sii local melting temperature, which depends linearly on salinity. \os %through the slope $\gamma_1$. 
This condition is supplemented  by a balance between heat and haline fluxes, which determine the melt rate following the widely adopted Stefan condition~\citep{middleton2021numerical,vreugdenhil2019stratification,gayen2016simulation}. 
 The conditions read in dimensionless form 
\begin{equation}
\left\lbrace
\begin{split}
%&z=0 \quad \text{(melting)}:\; \theta =-1 -\gamma_1  \sigma    \quad \text{and}   \quad \left. \dfrac{ \partial_z  \sigma}{(\sigma  + R_\rho)}\right|_{z=0} = \left.Le \so St \os  \partial_z \theta\right|_{z=0} \\
&z=0 \quad \text{(melting)}:\; \theta =-1 -\gamma_1  \sigma    \quad \text{and}   \quad \so \left. \dfrac{ \partial_z  \sigma}{(\sigma  + R_\rho)}\right. = \left.Le  St    \partial_z \theta\right. \os \\
&z=-1 \quad \text{(far field)}: \quad \theta = 0 \quad \text{and}   \quad \sigma = 0,
\end{split}
\right.
\label{eq:bc}
\end{equation}
with
\begin{equation}
\label{eq:numbers}
  \so R_\rho= \dfrac{S_\infty }{\Delta S } \text{ (density ratio)}, \os\; \gamma_1=- \lambda_1 \dfrac{\beta_T}{\beta_S} \text{ (freezing coupling),}   \;  \so St \os = \dfrac{C_p}{L_f} \Delta T \text{ (Stefan)}.
\end{equation}
\so The density ratio $R_\rho$ quantifies the balance between the stabilising effect of the far-field salinity and the destabilising temperature difference. \os
The parameter $\gamma_1$ represents a freezing coefficient coupling 
temperature and salt at the meting interface; 
the Stefan number $St$ characterises the relative importance of specific and latent heat, based on the initial temperature jump at the interface.
These boundary conditions are approximations that hold true only when the melting rate is slow. 
 Essentially, they substitute the mass flux resulting from the phase change with a diffusive flux, introducing a Lewis dependence in  Eq.~\eqref{eq:bc}.  This approach serves as an alternative to directly solving the Stefan problem, which, strictly speaking, requires explicit treatment of a moving interface---see, \emph{e.g.}, \citep{huppert1990fluid, wettlaufer2001stefan}. 
While phase-field methods can be employed to this end,  they prove computationally expensive for small melting rates, which produce a large time-scale separation  between the dynamics of phase change and that of the  fluid \citep{favier2019rayleigh, gastine2025rotating}.
 
Our numerical setup employs direct numerical simulations using the finite-volume code Oceananigans \citep{ramadhan2020oceananigans}. The computational domain is discretised on a grid of  $8192^2$ collocation points, uniform in the horizontal direction and refined vertically  near the top interface. \si For this resolution, the spectra of the scalar fields exhibits a clear decay at high wavenumbers (not shown), indicating that the smallest dynamically active scales are adequately resolved by the numerical grid. \is
Since Oceananigans operates with dimensional variables, all governing parameters are prescribed in physical units, for consistency with realistic seawater properties~\citep{couston2024turbulent, allende2024impact}.
 They  are summarised in Table~\ref{tab:therm_param}, and prescribe the free-fall velocity and time units as $U_0 \approx 0.098 \, m/s$ and $t_0 \approx 64 s \, $, respectively.

In all the simulations, we set $\Delta T = 2\mathrm{K}$. To  explicity satisfy the boundary conditions \eqref{eq:bc},
 we regularise the initial temperature jump by introducing a smooth temperature profile controlled by the  (small) parameter $\delta\ll L_x$---see Fig.\ref{fig:scheme}.  The parameter $\delta$ sets  the initial gradient $\propto \Delta T /\delta$, and offsets it down  from the upper boundary,  at a depth  $\propto \delta$. This smoothing ensures that the  temperature gradient vanishes at the top boundary, hence fulfilling the boundary  condition \eqref{eq:bc} imposed by the uniform salinity profile. 
To trigger the instability, the initial temperature profile of Fig.~\ref{fig:scheme} is perturbed by   a random noise
of small amplitude---\so see Appendix~\ref{sec:A} for additional details. \os
Our series of simulations vary the Lewis number from $1$ to $100$,  keeping the thermal diffusivity constant. 
The density ratio $R_\rho$ varies from  $1$ to $406$, tracking modifications in $S_\infty$.  
 The parameters in Table \ref{tab:therm_param} additionally prescribe the non-dimensional values:  $Pr=1$,  $St=2.4 \times 10^{-2} $ and $\gamma_1=2.8 \times 10 ^{-3}$, representative of polar ice–ocean conditions, though slightly idealised for numerical efficiency---in particular $Pr =O(10)$ in seawater.
\sii Additional simulations at $Pr=7$  (not shown) display similar flow behaviours.

\so
From an oceanographic perspective, please note that $R_\rho =O(10)$ corresponds to weakly saline conditions ($S_\infty \sim 1 g/kg$), which are representative of near-surface environments influenced by strong freshwater inputs. \sii By contrast, values $R_\rho \gtrsim 300$ corresponds to typical open-ocean salinities $S_\infty \gtrsim 30 g/kg$ encountered in polar oceans \citep{peralta2015seasonal}.
\so For these values, the linearised equation of state \eqref{eq:EoS} predicts that the density decreases monotonically with the temperature  and this is consistent with the observations---see, \emph{e.g.}, \citep{millero2010history,talley2011descriptive}. 
While the linearised equation of state is less accurate at lower salinities, it still provides a well-defined hydrodynamical framework, allowing for a systematic investigation of nonlinear melting dynamics and their sensitivity to key control parameters.
\sii In the present study, we therefore deliberately adopt a linear equation of state to retain a tractable framework and isolate the dominant physical mechanisms. The effects of a nonlinear equation of state, including the freshwater density anomaly, have been investigated in recent studies (e.g. \cite{yang2023ice,xu2025aspect}) and are beyond the scope of the present work.
\os

%######################################################################
%######################################################################
%######################################################################
% SECTION: RESULTS
\section{Results}
\label{sec:results}

\begin{figure}
 \center
%\begin{minipage}{0.67\textwidth}
\includegraphics[width=\textwidth, trim=0cm 0.cm 0cm 0cm, clip]{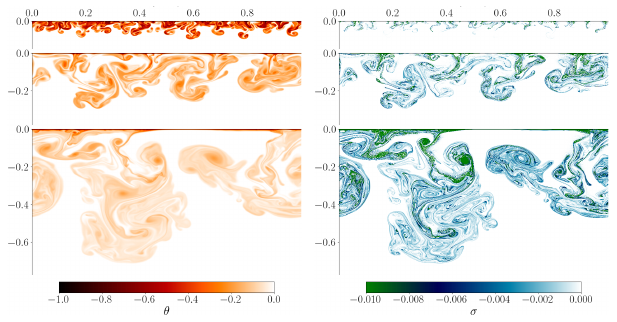}
%\end{minipage}
%\begin{minipage}{0.32\textwidth}
%\includegraphics[width=\textwidth, trim=0cm 1cm 6cm 0cm, clip]{Figs/evol4_add_sigma_Stop40.00_Tbot272.9_Le100_n8192}
%\includegraphics[width=\textwidth, trim=0cm 1cm 6cm 0cm, clip]{Figs/evol4_add_sigma_Stop0.10_Tbot275.2_Le100_n8192}
%\end{minipage}
%\caption{ Typical evolution of the temperature (a) and salinity fields (b) for a simulation with $Le = 100$ and $R_\rho = 406$. Panel (c) shows the temporal evolution of the  horizontally-averaged salinity profile for $R_\rho = 406$.  Panel (d) shows the results for $R_\rho = 1$.}
\caption{ \so Typical evolution of the temperature (left) and salinity fields (right) for a simulation with $Le = 100$ and $R_\rho = 406$. \os}

\label{fig:snapshots}
\setlength{\unitlength}{\columnwidth}
\begin{picture}(1,0)(0,0)
%\put(0.05,0.18){(a)}
%\put(0.39,0.18){(b)}
%\put(0.73,0.34){(c)}
%\put(0.73,0.15){(d)}

\put(0.99,0.55){ $t=5$}
\put(0.99,0.45){$t=10$}
\put(0.99,0.22){$t=20$}

\end{picture}
\end{figure}

\begin{figure}
\center
\includegraphics[width=0.49\textwidth, trim=0cm 0cm 0cm 0cm, clip]{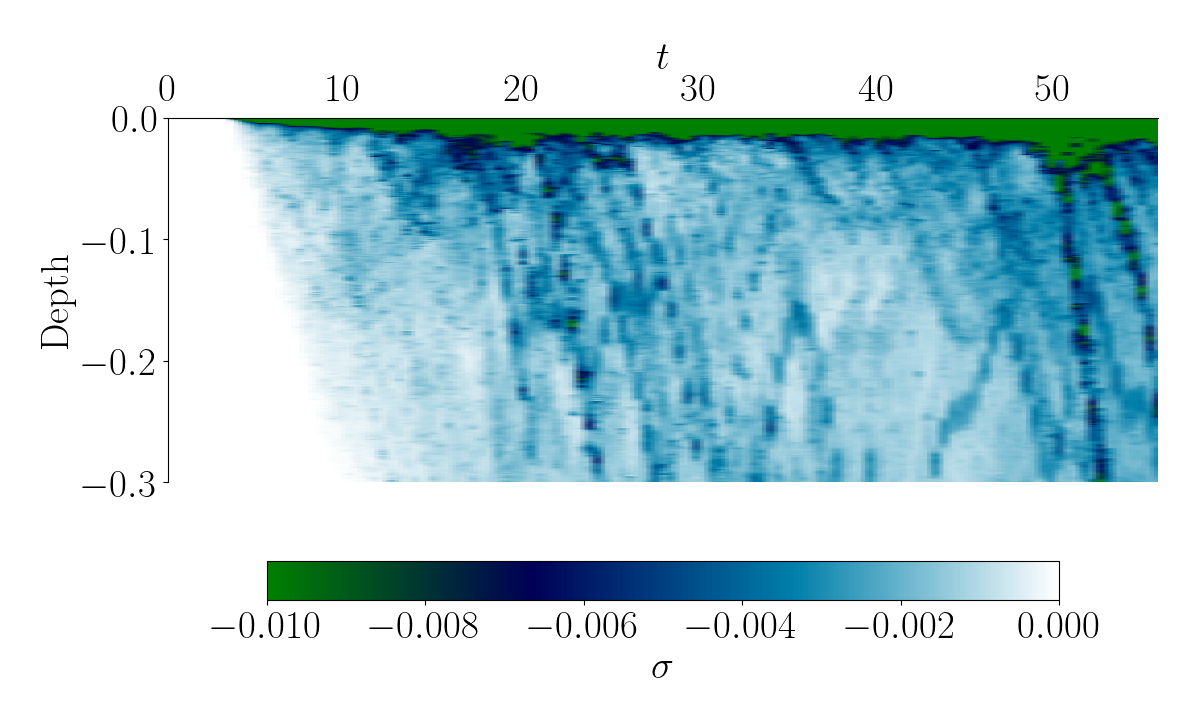}
\includegraphics[width=0.49\textwidth, trim=0cm 0cm 0cm 0cm, clip]{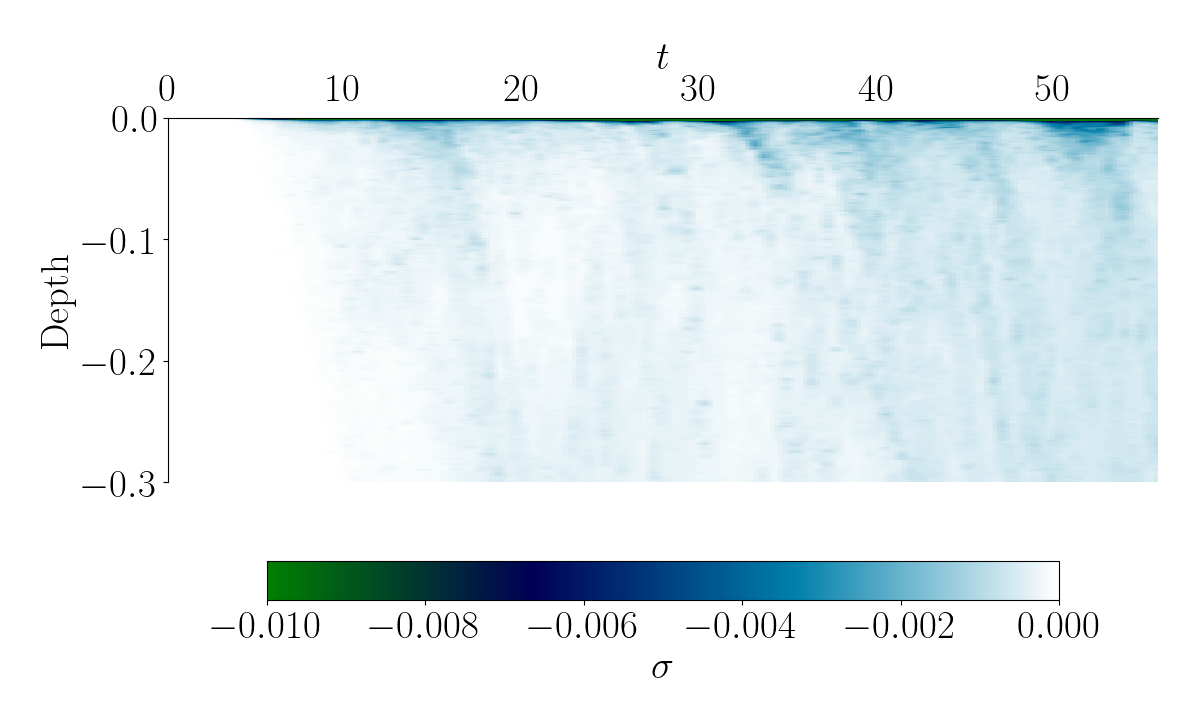}
\caption{\so Left: Temporal evolution of the  horizontally-averaged salinity profile for $Le = 100$ and  $R_\rho = 406$. Right: Same but for  for $Le = 100$ and  $R_\rho = 1$. \os}
\label{fig:evolution}
\end{figure}

In all simulations,  the fluid evolves through three main phases.  \so This is illustrated in Fig.~\ref{fig:snapshots}, showing representative temperature and salinity snapshots.  \os The system initially develops a Rayleigh\textendash Taylor-like instability, which arises from  the initially unstable buoyancy jump away from the interface \citep{drazin2004hydrodynamic, boffetta2017incompressible}.
This is followed by the interaction of the flow with the upper boundary and the onset of melting. Eventually, a mixing layer develops near the upper boundary and gradually deepens, with  cold and fresh  plumes extending downward into the domain. Note that contrary to previous studies with similar configurations~\citep{guo2025effects}, we here focus on the transient dynamics before the turbulent layer reaches the bottom of the numerical domain and will not comment on the subsequent stationary (or quasi-stationary) states. 
 \so Varying $Le$ and $R_\rho$, the main %qualitative 
difference is seen not in the bulk but at the melting interface. Increasing $R_\rho$, the top interfacial layer of fresh water %near the top 
gets thicker, as seen from %the dynamics of 
the horizontally-averaged salinity profiles in Fig.~\ref{fig:evolution}. \os No obvious difference is seen for the temperature, neither in the bulk nor at the interface.
 These observations suggests the following phenomenology:
Salty environments favour strong stratification between the relatively fresh meltwater layer and the ambient ocean  leading to a stable diffusive layer, while low density ratios favour mixing close to the upper interface.

%######################################################################
%######################################################################
%######################################################################
% SECTION: DRIVING BOUNDARY CONDITIONS
\subsection{Fluxes and temperature at the top interface}
\begin{figure}
 \center
\includegraphics[width=\textwidth, trim=0cm 1.9cm 0cm 0cm, clip]{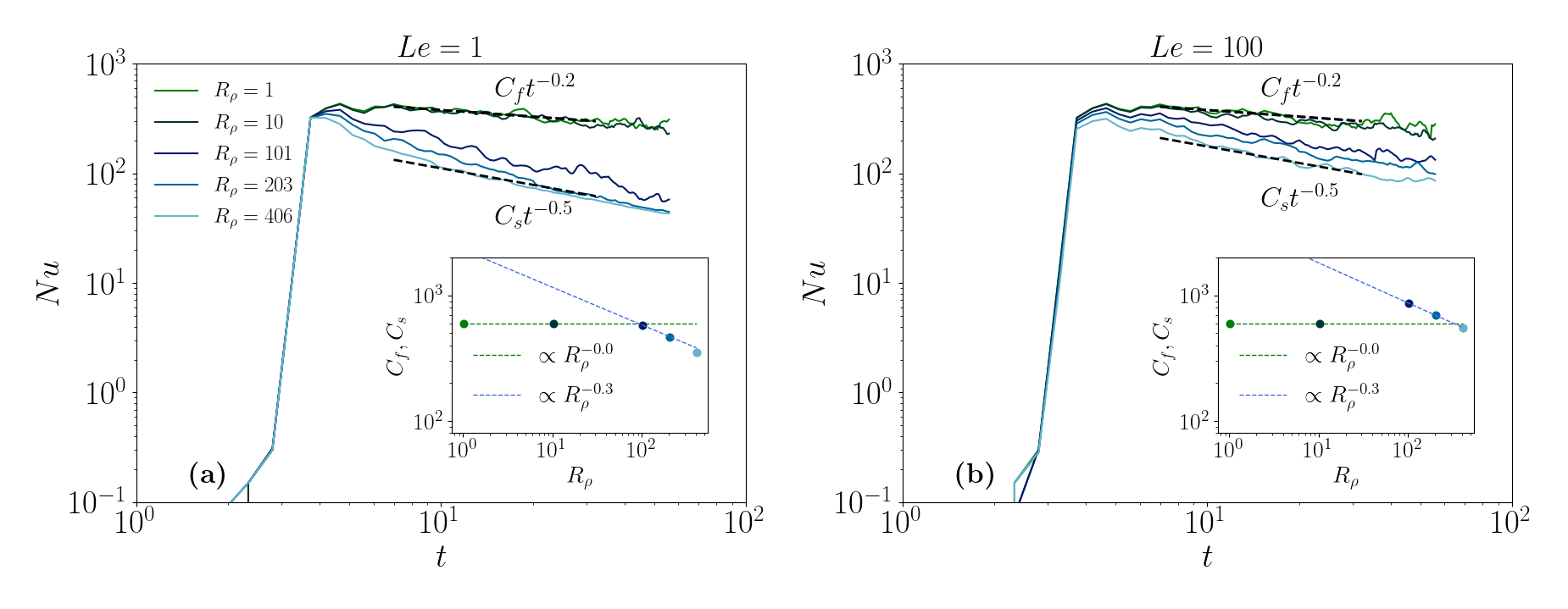}
\includegraphics[width=\textwidth, trim=0cm 0cm 0cm 0cm, clip]{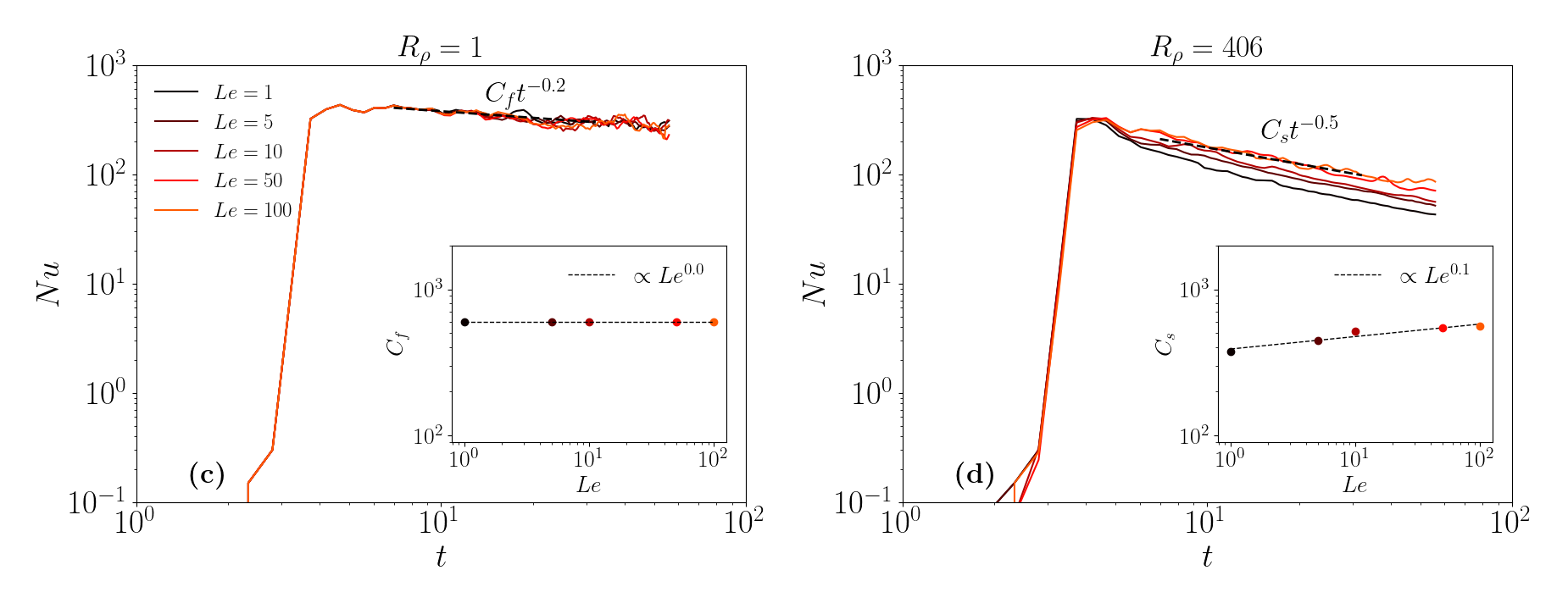}
\caption{Time evolution of the Nusselt number $Nu$,  varying $R_\rho$ for fixed $Le$ (top) and $Le$ for fixed $R_\rho$  (bottom). \si In all four panels, the insets display  the prefactors $C_s, C_f$ (see text) averaging over times $ t \gtrsim 10$.}
\label{fig:Nu}
\end{figure}

To quantify the melt-rate, we introduce the Nusselt number,
\begin{equation}
Nu = -   \left\langle \partial_{ z} \theta \right\rangle_{z=0},
\label{eq:nu}
\end{equation}
with $\left\langle\cdot \right\rangle$  denoting horizontal averaging.
Analogous to Rayleigh\textendash Bénard (RB) convection, the Nusselt number measures the  thermal flux at the top boundary in units of the average temperature gradient $\Delta T/L_x$---recall that $L_x =1$ in our length units. It  is directly proportional  to the melt rate, $Nu \propto Le \, \so St \os \left. \partial_z \theta \right|_{z=0}$. With the conventional minus sign in \eqref{eq:nu},  \so the regime $Nu > 0$   reflects melting  (this is the case under study) while $Nu<0$ would indicate freezing. \os
Unlike  RB convection, $Nu$ decays over time. Its algebraic rate of decay  exhibits two distinct asymptotic regimes, highlighting a strong dependence on the density ratio $R_\rho$ and only a weak dependence on $Le$---see Fig.~\ref{fig:Nu}.
\is
In fresh environments, \emph{i.e}, for small values of  $R_\rho \lesssim 10$, the Nusselt number decays only mildly  as $Nu\sim C_f t^{-0.2}$, where both the  exponent and the prefactor are insensitive to the Lewis number.
In salty environments, \emph{i.e}, for large values of  $R_\rho \gtrsim 10$, we observe a diffusive scaling  $Nu\sim C_s t^{-0.5}$,
with the prefactor following the apparent scaling law $C_s \sim Le^{a} R_\rho^{b}$. Our simulations indicate $ a\simeq 0.1$ and $b\simeq -0.3$,
meaning that the Lewis and density ratio have opposite effects, respectively enhancing and slowing melting  in this apparent diffusive regime. 

\begin{figure}
 \center
\includegraphics[width=\textwidth, trim=0cm 1.9cm 0cm 0cm, clip]{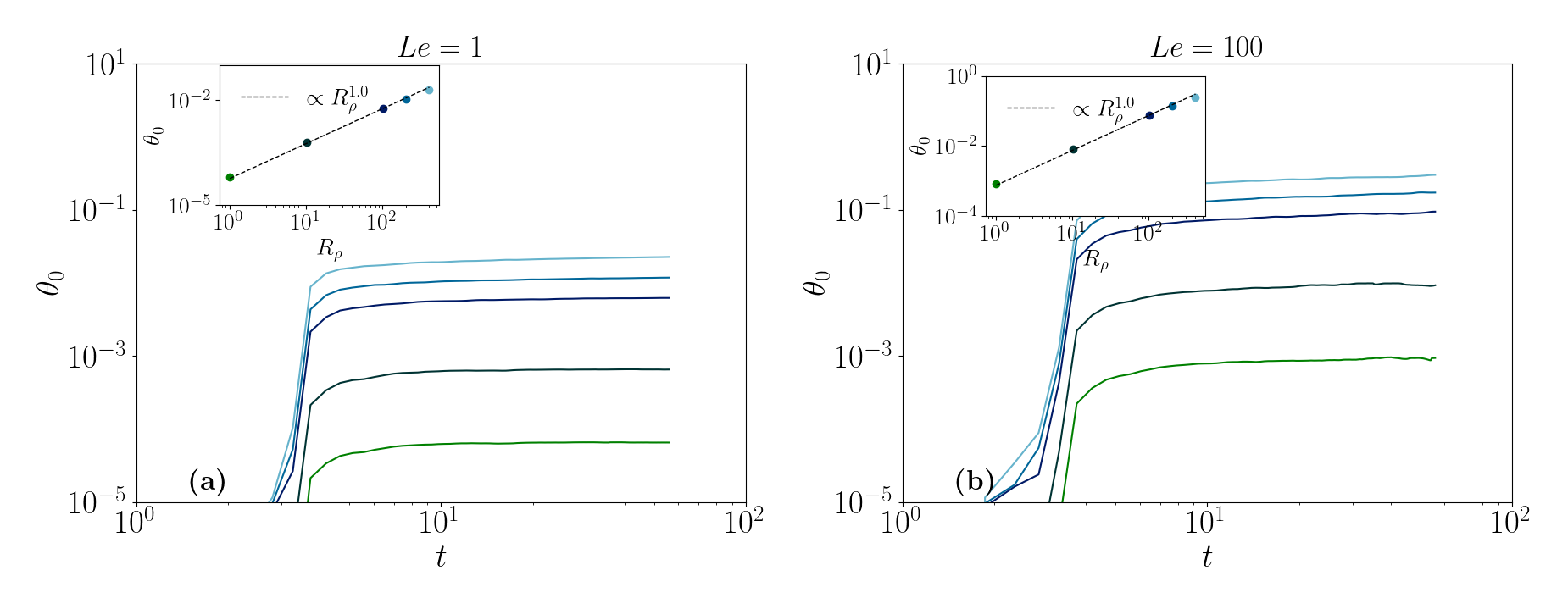}
\includegraphics[width=\textwidth, trim=0cm 0cm 0cm 0cm, clip]{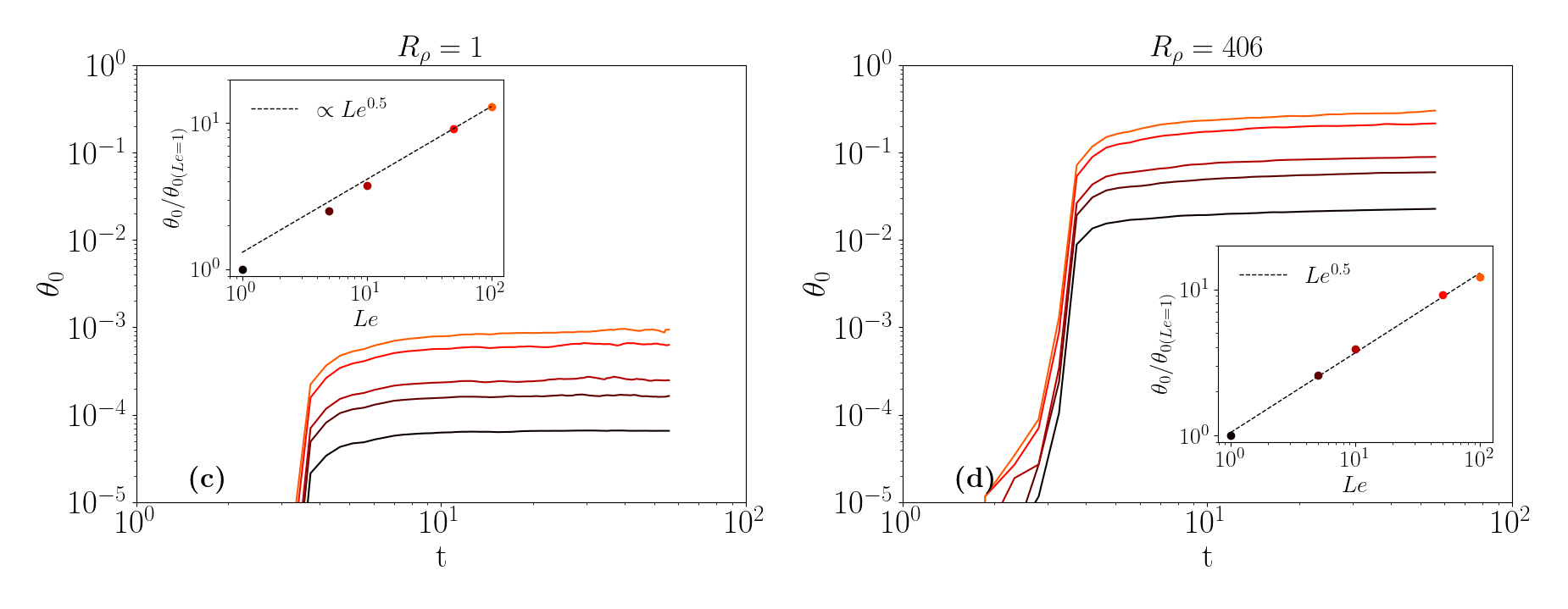}
\caption{Time evolution of the excess temperature $\theta_0$,  varying $R_\rho$ at fixed $Le$ (top) and $Le$ at fixed $ R_\rho$  (bottom). \si In all four  panels, the insets  display the dependence on either $R_\rho$ or $Le$ averaging over times $ t \gtrsim 10$.}
\label{fig:thetatop}
\end{figure}

The salinity at the boundary modifies the freezing temperature  at the interface. Fig.~\ref{fig:thetatop} shows the evolution with time of  the  excess temperature  (\emph{i.e.}, above the freezing temperature based on the far-field salinity)
\begin{equation}
	\theta_0 := 1+\left\langle \theta \right\rangle_{x,z=0}  =-\gamma_1 \sigma_0,
\end{equation}
where $\sigma_0=\left\langle \sigma \right\rangle_{x,z=0}$.
The second equality indicates that $\theta_0$ is proportional to the horizontally averaged  fluctuation in salinity---see Eq.\eqref{eq:bc}. 
 Positive fluctations $\theta_0 \ge 0$ are associated with fresh water   at the interface, $\sigma_0 \le 0$.
Unlike the Nusselt number, no clear dichotomy emerges. 
In all the simulations, the top temperature  quickly reaches a \so quasi-stationary \os value, yielding  the  scaling behavior 	$\theta_0 \propto Le^{0.5} R_\rho$. 
\so Theoretically, this scaling reflects a diffusive dynamics near the melting boundary---see Appendix \ref{sec:B}. \so
It is consistent with the intuitive idea that  saltier ambient water  favors a  comparatively fresher interfacial boundary layer near the top interface. \os

%######################################################################
%######################################################################
%######################################################################
% NEW SECTIONS FOR REVISION: MELTWATER TRANSPORT
\subsection{Meltwater transport: from diffusion to convection}
\so Our setup employs a regularization that offsets the initial temperature jump  between  the ice and the underlying bulk fluid  by a depth $\delta\ll 1$. \is
This triggers a classical Rayleigh\textendash Taylor turbulent mixing layer.  Without an upper  boundary, the layer would  grow  quadratically  $L(t)\sim  t^2$  under a critical balance between buoyancy and nonlinear terms---this is  the self-similar Bolgiano\textendash Obukhov  scenario \citep{chertkov2003phenomenology,boffetta2017incompressible}. 
\so The interaction with the top interface, however, disrupts this classical regime by creating a (meltwater) boundary layer that couples the melting interface to the turbulent mixing layer. 
\begin{figure}
\center
\begin{picture}(1,0)(0,0)
\put(-110,1){ Temperature}
\put(90,1){Salinity}
\end{picture}
\includegraphics[width=\textwidth, trim=0cm 1cm 0cm 0cm, clip]{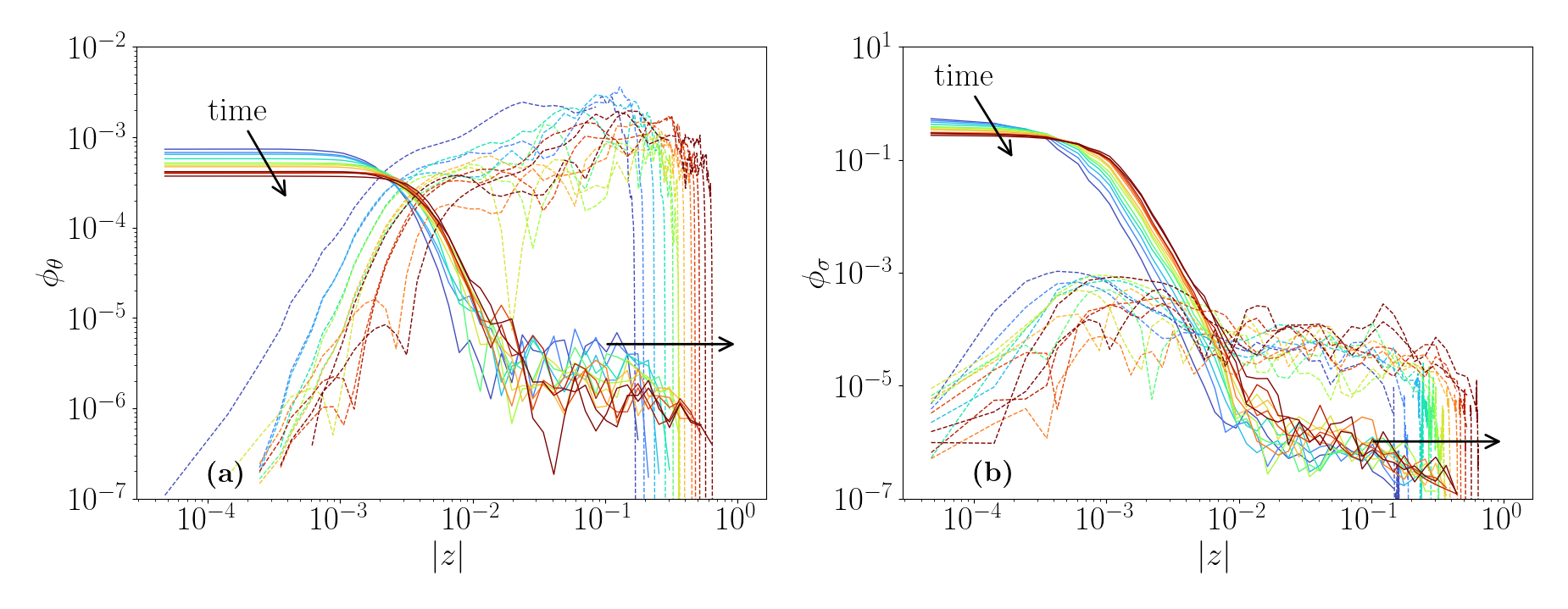}
\includegraphics[width=\textwidth, trim=0cm 1cm 0cm 0cm, clip]{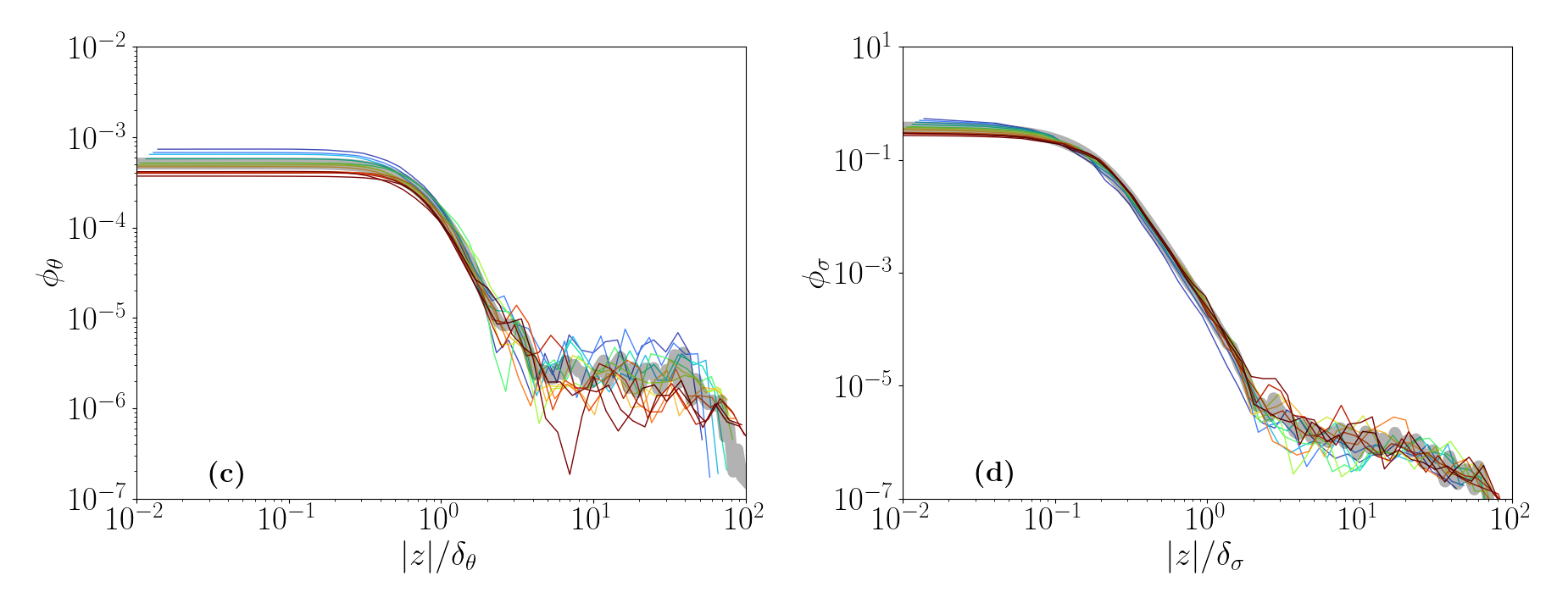}
\includegraphics[width=\textwidth, trim=0cm 0cm 0cm 0cm, clip]{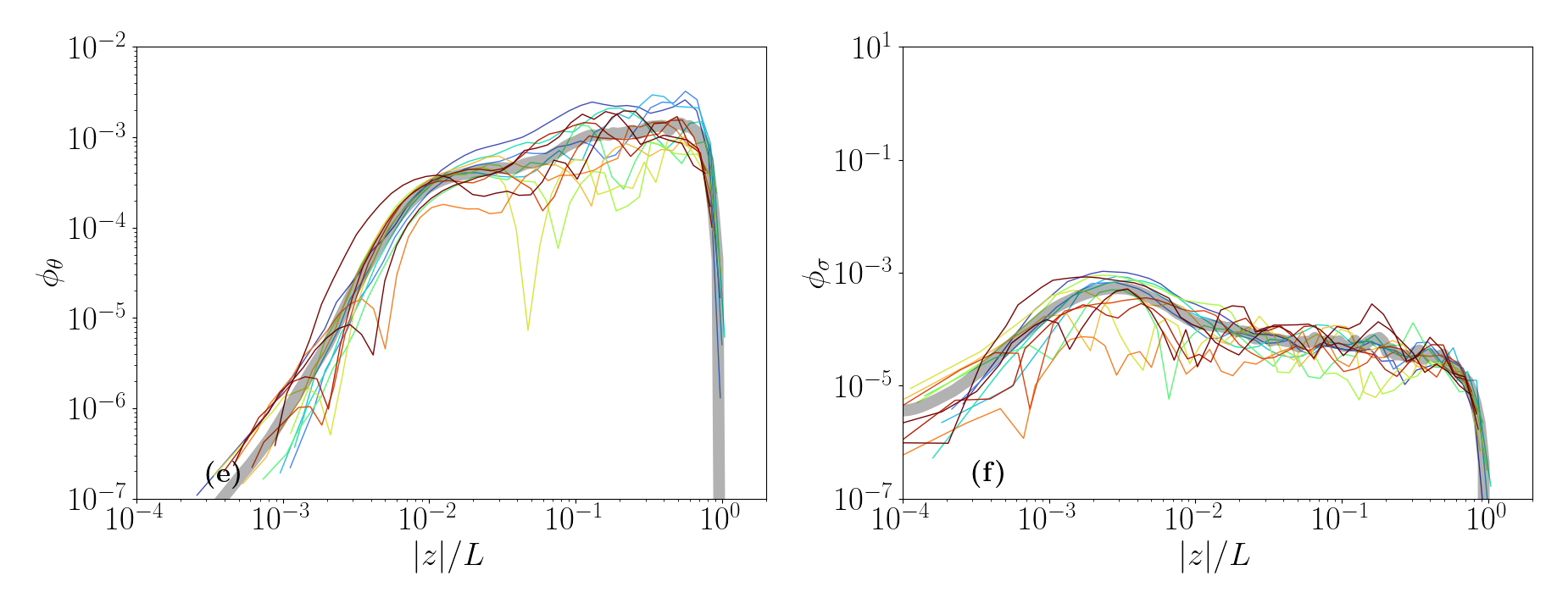}
\caption{Top: Vertical fluxes  at successive times for the case  $Le = 100$ and $R_\rho = 406$,  decomposed into their diffusive (solid lines) or convective (dashed lines) contributions. 
Middle:  Diffusive contributions rescaled using the diffusive scaling $\delta_\chi(t) \propto t^{1/2}$. Bottom:  Convective contributions  rescaled using the superdiffusive scaling $L(t) \propto t^{4/3}$.
Panels (a,c,e) are for the thermal fluxes. Panels (b,d,f) are for the solutal fluxes.}
\label{fig:differentialfluxes} 
\end{figure}
To characterise the exchange between the  boundary layer and the turbulent layer, we examine the vertical fluxes of temperature and salinity. After horizontal averaging, each flux can be decomposed into convective and  diffusive contributions as 
\begin{equation}
	\phi_\chi=  \phi_\chi^c + \phi_\chi^d,\quad \phi_\chi^c = \left\langle \chi u_z  \right\rangle,\quad \phi_\chi^d = -\lambda_\chi^{-1} Pr^{-1}Re^{-1} \left\langle  \partial_z \chi \right\rangle, \quad \chi \equiv \theta,\sigma
\end{equation}
with $\lambda_\theta = 1$ and  $\lambda_\sigma=Le$.
Figure~\ref{fig:differentialfluxes} illustrates the generic behaviour using the most saline environment $R_\rho =406$ at $Le=100$, highlighting two propagation rates.
\si
At any time during the growth,  the convective fluxes $\phi_\theta^c,\phi_\sigma^c$  fluctuate in $z$ about a nearly  constant value before dropping sharply. These  drops become synchronised  from $t\gtrsim 2$, defining the extent of the  convective mixing layer. 
\is
%Fix a  time: The convective fluxes $\phi_\theta^c,\phi_\sigma^c$  fluctuate in $z$ about a nearly  constant value before dropping sharply. These  drops become synchronised  from $t\gtrsim 2$, defining the extent of the  convective mixing layer. \sii 
The layer  propagates superdiffusively and reaches the bottom interface at $t\simeq 27$  as  \so
\begin{equation}
	L(t) \simeq C \, t^{\beta},\quad  C \approx  0.015,\quad \beta\simeq 1.33
	\label{eq:ml43}.
\end{equation}
The diffusive contributions $\phi_\theta^d,\phi_\sigma^d$ evolve on slower timescales, following the (diffusive) scalings
\begin{equation}
%	\delta_\theta(t)  = Re^{-1/2}Pr^{-1/2} t^{1/2},\quad \delta_{\sigma(t)}=Le^{-1/2} Re^{-1/2}Pr^{-1/2} t^{1/2};
	\delta_\chi(t)  = \lambda_\chi^{-1/2}Re^{-1/2}Pr^{-1/2} t^{1/2},\quad \chi\equiv\theta,\sigma%\quad \delta_{\sigma(t)}=Le^{-1/2} Re^{-1/2}Pr^{-1/2} t^{1/2};
	\label{eq:diffscaling}
\end{equation}
see Appendix \ref{sec:A}. 
In Fig.~\ref{fig:differentialfluxes},  this differential propagation is signaled by the collapse of the convective and diffusive  fluxes shown in the panels c,d,e and f. These collapses are obtained by rescaling \sii lengths \so with  the superdiffusive scaling \eqref{eq:ml43}  and the diffusive scaling \eqref{eq:diffscaling}, respectively.
At each time, the crossovers between  diffusive and convective contributions define the effective boundary layer thicknesses $\propto  \delta_\theta,\delta_\sigma$  for the solutal and thermal field, respectively. Within this layer, the transport of meltwater is diffusive and hence sensitive to double-diffusive effects. Below this layer,  solutal and thermal transport is  convective, but in a non-trivial way. While this convection originates from the  nonlinear stages of the initial RT instability, it does not propagate following the critical-balance scaling $\propto t^2$. Instead, the superdiffusive rate \eqref{eq:ml43} reflects the presence of a boundary at $z=0$.

%%%%%%%%%%%%%%%%%%%%%%%%%%%%%%%%%%%%%%%%%%%%%%
%%%%%%%%%%%%%%%%%%%%%%%%%%%%%%%%%%%%%%%%%%%%%%
%%%%%%%%%%%%%%%%%%%%%%%%%%%%%%%%%%%%%%%%%%%%%%
%%%%% CONSTANT FLUX SCENARIO
\subsection{Constant-in-space heat flux scenario for  the superdiffusive growth $\propto t^{4/3}$}
\label{sec:constantflux}
The collapse in  Fig.~\ref{fig:differentialfluxes} with the superdiffusive rates \eqref{eq:ml43} suggests a constant-flux propagation scenario through the domain. This can arise  if the convective thermal and solutal fluxes evolve self-similarly as
\begin{equation}
		 \phi_\chi^c (z,t)  \sim  L^{-\alpha}\Phi_{\chi}\left(  \dfrac{|z|}{L(t)}\right) + \text{fluctuations},\quad\chi \equiv \theta,\sigma,
\label{eq:ss_fluxes}
\end{equation}
involving similarity flux profiles $\Phi_\theta,\Phi_\sigma$.
\begin{figure}
 \center
\includegraphics[width=\textwidth]{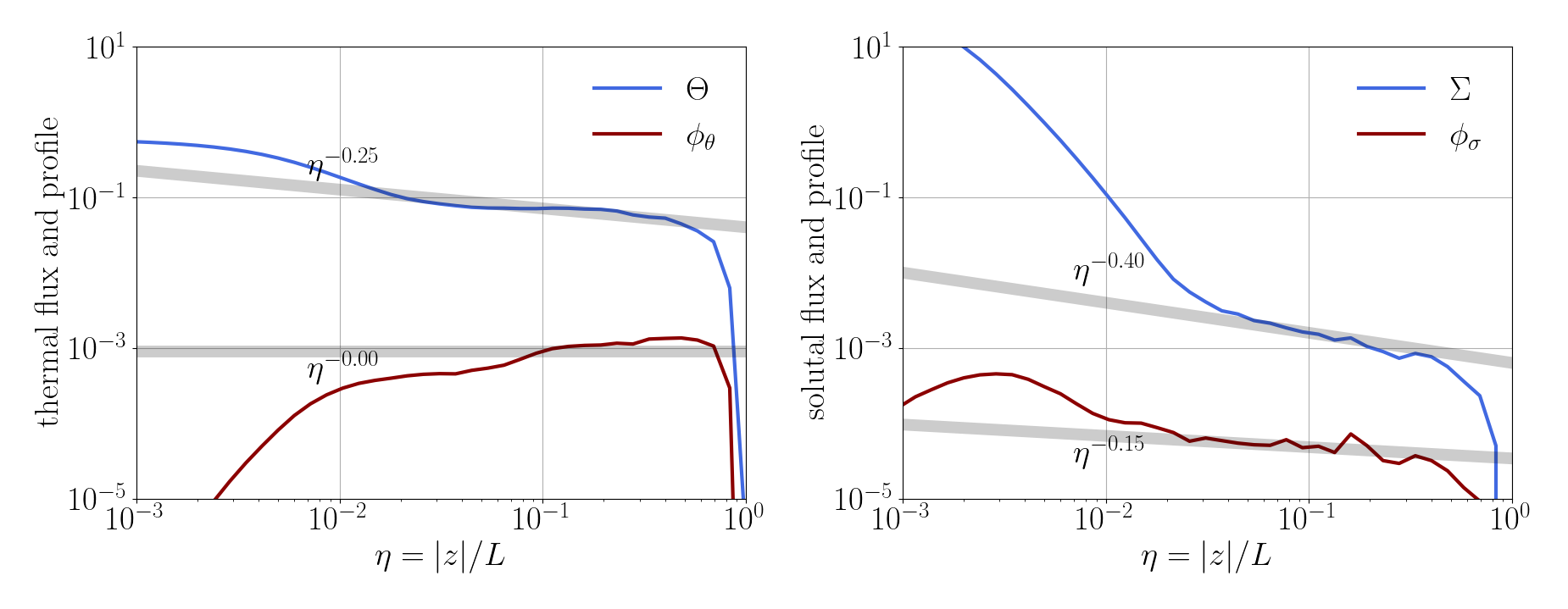}
\caption{ Similarity profiles obtained from the  scalar fields and  fluxes  for $Le = 100$ and $R_\rho = 406$. Left: Temperature, with  the exponents $\alpha=0,x=0.25$. 
Right: \si Salinity with the exponents $\alpha=0.15,x=0.4$. \is } 
\label{fig:Similarity} 
\end{figure}
The exponent $\alpha$ prescribes the decay rate $\propto (|z|/L)^{-\alpha}$ of the convective flux across the mixing layer.
For example, setting $\alpha=0$ gives a scenario where  the convective fluxes are approximately uniform across the mixing layer.
Regardless of the exponent, the horizontally averaged fields must also be self-similar within the mixing layer,
\begin{equation}
		 \left\langle \theta\right\rangle (z,t)  \sim L^{-x} \Theta\left(  \dfrac{|z|}{L(t)}\right) + \text{fluctuations},\quad
		 \left\langle \sigma\right\rangle (z,t)  \sim L^{-x} \Sigma\left(  \dfrac{|z|}{L(t)}\right) + \text{fluctuations},\quad
\label{eq:ss_p}
\end{equation}
for some exponent $x \ge 0$. This form implies the algebraic behavior $\propto (|z|/L)^{-x}$ far away from the front but within the mixing layer, \emph{i.e.} $\delta_\theta,\delta_\sigma \ll |z| \ll L$---see, \emph{e.g.}, \citep{barenblatt1996scaling,campolina2025non}. In turn, this prescribes the behaviors for the similarity profiles
\begin{equation}
	\Phi_{\chi}(\eta) \sim |\eta|^{-\alpha},\quad \Sigma(\eta),\; \Theta(\eta) \sim |\eta|^{-x}\quad |\eta|\ll 1.
\end{equation}
Assuming an algebraic growth $L\propto t^\beta$, the dimensional balance between the transport terms $\partial_z \phi^c_\theta$ and $\partial_t \left\langle \theta\right\rangle $  prescribes that the three exponents $\beta,x,\alpha$ satisfy
\begin{equation}
	\beta(1+\alpha-x)=1
\label{eq:compatibility}
\end{equation}
Provided that $x<1$, then  the temperature  field evolves  as 
\begin{equation}
	\dfrac{d}{dt}\int_{-\infty}^0 dz \; \theta(z,t) = 
	\left\lbrace
		\begin{split}
			& (PrRe)^{-1} Nu(t) \\
			& \beta(1- x) t^{\beta-\beta x -1} \int_{0}^1\Theta(\eta) d\eta  \propto t^{-\alpha\beta}
		\end{split};
	\right. 
\label{eq:compatibility2}
\end{equation}
where the  first and second lines in the rhs are obtained from the boundary flux and from the self-similar scaling, respectively.

\si Combined with  the growth exponent $\beta=1.33$ from Eq. \eqref{eq:ml43}, the compatibility Eq.~\eqref{eq:compatibility}
  prescribes in particular  $x-\alpha \simeq 0.25$ leaving only one exponent to be determined, say $\alpha$. In turn, Eq.~\eqref{eq:compatibility2} relates the exponent $\alpha$ to the decay of Nusselt.
\si
The Nusselt decay $Nu \propto t^{-0.2}$ reported in Fig.~\ref{fig:Nu} suggest $\alpha\beta \simeq 0.2$, implying $\alpha \simeq 0.15$ and $x \simeq 0.4$. This scenario prescribes  that the heat flux decreases with depth  across the mixing layer $\phi_\chi \propto |z|^{-0.15}$.
%On the other hand, let us point out that  
On the other hand, Fig.~\ref{fig:Nu} could possibly reflect constant behaviour $Nu \propto t^{0}$---only polluted by finite-size effects. This constant-flux scenario implies  $\alpha=0$ and $x\simeq 0.25$.
Both scenarios can be examined directly. To compute the universal profiles $\Theta$, $\Phi_\theta$, etc, we suppress the fluctuations in Eq.~\eqref{eq:ss_fluxes} by first rescaling $ z \to z/L(t)$, multiplying by the suitable power of $L$ and then  temporally averaging. The results are shown  in Fig.~\ref{fig:Similarity}.
Inspection of the thermal flux supports the exponent $\alpha=0$, corresponding to an approximately  constant flux across the domain and implying %Together with  $\beta\simeq 1.33$, this  constant-flux scenario  implies 
%$x\simeq 0.25$ from Eq.~\eqref{eq:compatibility}. 
the  scaling $\eta^{-0.25}$ for thermal layer away from the front.
%Figure~\ref{fig:Similarity} reveals another feature. While the growths of the thermal and solutal layer are synchronised with the rate \eqref{eq:ml43}, the similarity profiles differ: t
The  solutal layer is  rather characterised  by  the exponents $\alpha\simeq 0.15$ and $x\simeq 0.4$.  The difference between  thermal and solutal similarity may reflect the asymmetric roles of temperature and salinity, which respectively provide destabilising and stabilising contributions to the initial density profile. \is

\so

\subsection{Universality of the convective regime}
\label{sec:universality}
 \begin{figure}
 \center
\includegraphics[width=\textwidth, trim=0cm 0cm 0cm 0cm, clip]{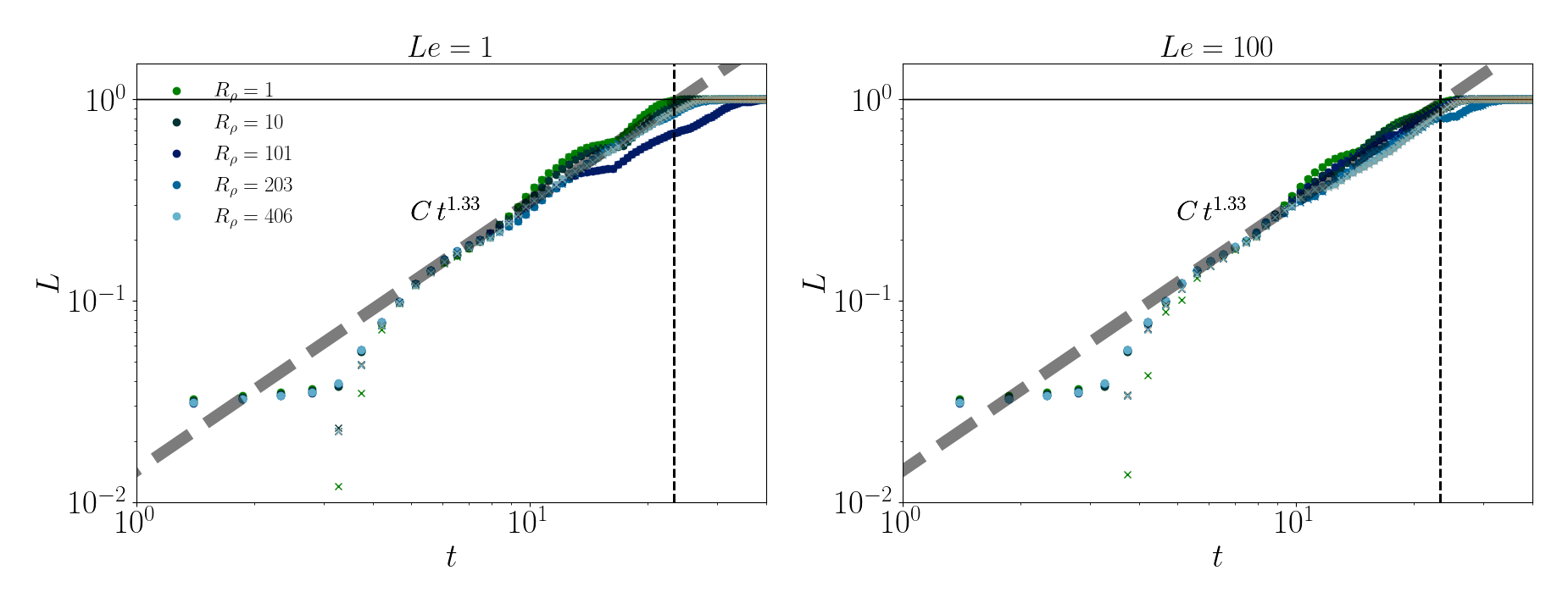}
\caption{Left: Time evolution of the thermal (o) and solutal (x) mixing lengths at $Le=1$ for various density ratios.  The vertical dashed line marks the time at which the fronts reach the bottom boundary.  Right: Same but for $Le=100$. 
} 
\label{fig:theta_sigma_mlevol}
\end{figure}
The convective regime is largely insensitive to variations of the density ratio and the Lewis number.
\si To provide a systematic estimate of the convective mixing layer, we compute $L(t)$ by thresholding the heat   flux at a small cutoff value $\epsilon =10^{-12}$, explicitly defining 
\begin{equation}
L(t) = \sup \left\lbrace |z|, \phi_\theta(t) \ge 10^{-12} \right\rbrace.
\end{equation}
 \is Figure \ref{fig:theta_sigma_mlevol} gathers the data, revealing the universality of the superdiffusive rate \eqref{eq:ml43} across the range of $Le$ and $R_\rho$ investigated. This universality is compatible with the similarity scenario sketched in \S~\ref{sec:constantflux}: the mixing layer growth is driven by a constant-flux similarity solution, associated to the inviscid dynamics within the bulk. The latter is obtained by formally setting $Re \to \infty$ in \eqref{eq:ns-boussinesq}. This scenario explains why the constant $C_f$ measured in Fig.~\ref{fig:Nu} is independent both from $Le$ and $R_\rho$.

\sii Double-diffusive effects are  present only close to the upper interface. \so They  regulate the exchanges between the boundary meltwater layer and the turbulent mixing layer.  We quantify  these exchanges through the efficiency ratios
\begin{equation}
	\Gamma_\chi(t) = \dfrac{ |\Phi^c_\chi(\eta=0.5)|}{|\phi^d_\chi(z=0)|},\quad \chi\equiv \theta,\sigma,
%	\Gamma_\chi(t) = \dfrac{|\phi^d_\chi(z=0)|}{ |\Phi^c_\chi(\eta=0.5)|},\quad \chi\equiv \theta,\sigma,
\end{equation}
 estimating the relative strength of  the convective flux---estimated halfway towards the front---
 compared with the diffusive flux  at the interface.
\begin{figure}
 \center
\includegraphics[width=\textwidth, trim=0cm 0cm 0cm 0.8cm, clip]{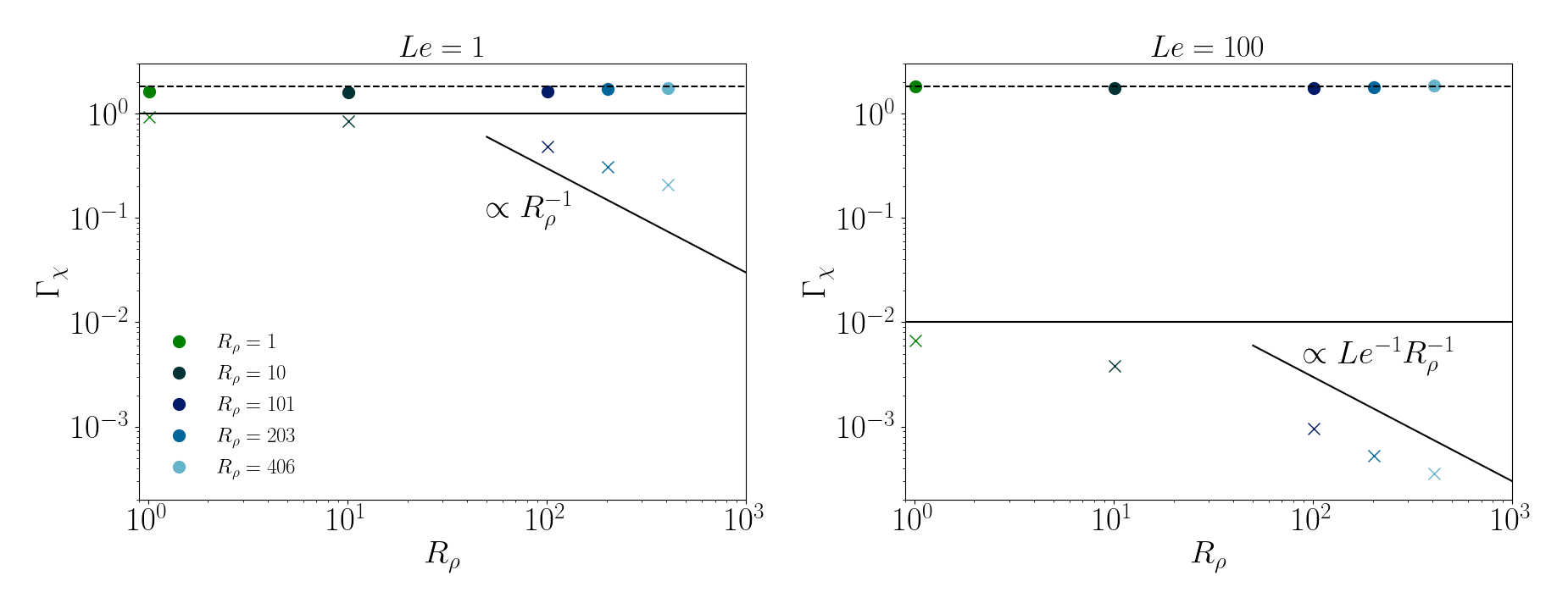}
\caption{\so Thermal flux ratio $\Gamma_\theta$  (circles) and solutal  flux ratio $\Gamma_\sigma$  (crosses) for different  density ratios. 
 Left: $Le = 1$. Right: $Le=100$.
The horizontal lines indicate  $\Gamma_\theta =1.8$ (dashed) and $\Gamma_\sigma =Le^{-1}$ (solid).  The indicative scalings $\propto Le^{-1}R_\rho^{-1}$ share a common prefactor  $ \simeq 30$.
\os} 
\label{fig:Fluxratios} 
\end{figure}
Figure~\ref{fig:Fluxratios} shows the result at a  time $t = 10$, when the turbulent layer is developed but occupies only $1/5$ of the box---still far from the bottom boundary.  
For the  thermal fluxes, the efficiency ratio $\simeq 1.8$ is largely independent of the density ratio and Lewis number. %There is a slight enhancement of diffusive flux as $R_\rho$ decreases below 10, marking a transition between relatively saline and relatively fresh environments.
A transition at $R_\rho \lesssim 10$ between relatively saline and relatively fresh environments is  apparent for  the solutal efficiency. %These are constant within the mixing layer and peaked at the melting interface, indicating formation of a  fresh meltwater layer that decouples the  solid interface from the turbulent bulk.  We measure the amplitude of the  salt-layer induced decoupling  by comparing the  flux of fresh water at the interface, dominated by diffusion,  with that at  $z=L/2$ in the midst of the turbulent layer--- dominated by convection. 
For a fixed Lewis, the solutal efficienty varies midly  in the  fresh case,  $R_\rho \lesssim 10$, and decreases  algebraically  $\propto R_\rho^{-1}$ in the salty case, $R_\rho \gtrsim 10$. This transition is compatible with that observed in Fig.~\ref{fig:Nu} for the time evolution of the  Nusselt number, marking the onset of the  regime of  diffusive decay $Nu \propto t^{-0.5}$ and  the formation of the (fresh) diffusive boundary layer near the top interface. This layer does not impact the  statistics of turbulent mixing within the bulk,   but regulates the amount of salt deficit injected into it.  
The efficiency decreases with Lewis; Asymptotic estimates relying on 
%The growth $ \Gamma_\sigma \propto Le R_\rho$ stems from 
the diffusive nature of the boundary layer dynamics predict $\propto Le^{-1}R_\rho^{-1}$. Such estimates are obtained from the diffusive phenomenology described in Appendix~\ref{sec:B}---in particular the  gradient estimate \eqref{eq:Derivative} $\partial_z \sigma \propto \sigma_0 Le^{1/2}/\sqrt{t}$  and the scaling estimate \eqref{eq:scaling} $\sigma_0 \propto R_\rho Le^{1/2}$.   This scaling law is shown in Fig.~\ref{fig:Fluxratios} and is compatible with our observations. It implies that only a small fraction $\propto Le^{-1}R_\rho^{-1}$ of the input solutal flux permeates through the interfacial layer to reach the convective layer.

\os

%######################################################################
%######################################################################
%######################################################################
% SECTION: What defines the mixing length?
\subsection{Comments on  mixing lengths}
 At high $R_\rho$, the physics involves an interplay between turbulent and diffusive flows,  making the interpretation of  mixing lengths strongly dependent on their practical evaluation. The mixing length $L(t)$ used in the previous section tracks the turbulent front, and is hence unaffected by  the fresh boundary layer.
%\si One may measure this length  by thresholding either the  thermal or the solutal profile at a very small value $L_\theta \leq 10^{-12}$ and $L_\sigma \leq 10^{-12}$. \is 
It  emphasises the role of thermally-induced convection and is largely independent of $Le$ and double-diffusive effects within the bulk. 

\begin{figure}
 \center
\includegraphics[width=\textwidth, trim=0cm 0cm 0cm 0.4cm, clip]{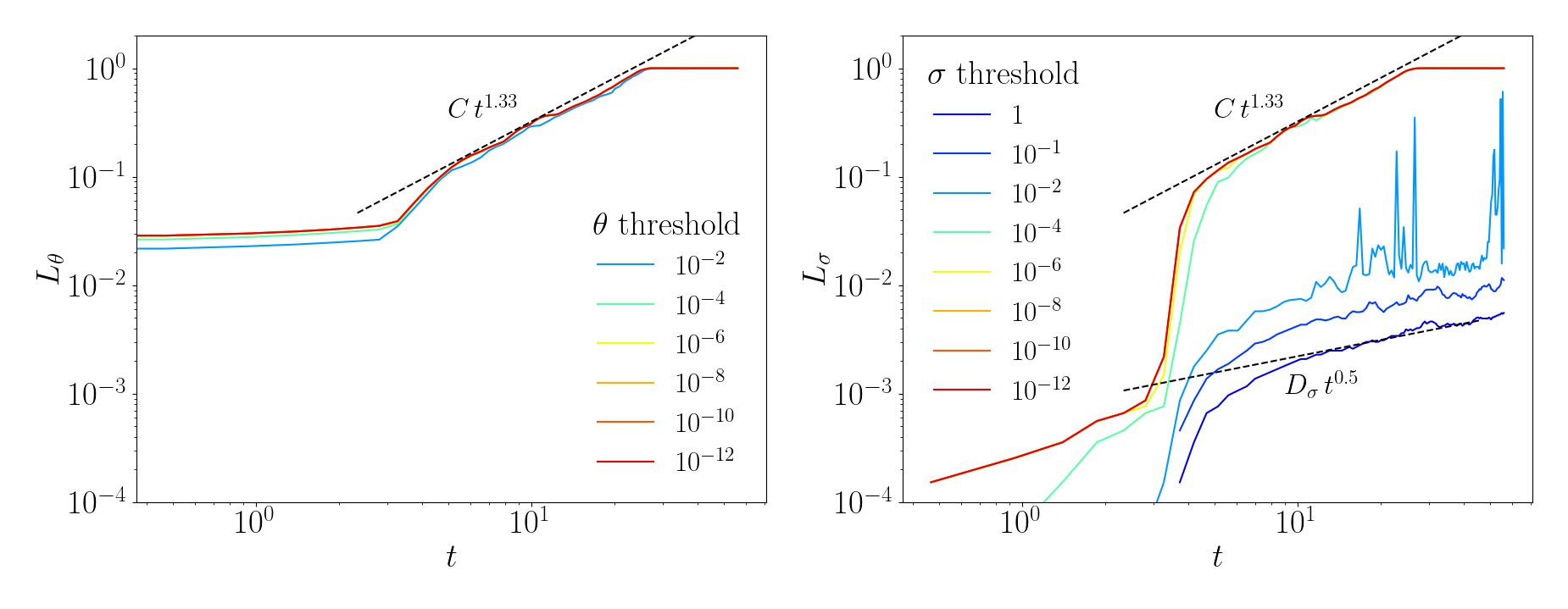}
\caption{\so Temporal evolution of the mixing length evaluated  at $Le=100$ and $R_\rho=406$ using different thresholds. Left: Thermal mixing length. Left: solutal mixing length. \os}
\label{fig:ml_theta_sigma}
\end{figure}

Among various options, the temperature and salinity mixing layers could be defined from  different threshold values \sii on the scalar fields themselves, \os as  illustrated in Fig.~\ref{fig:ml_theta_sigma}. We observe that the temperature mixing layer exhibits the universal super-diffusive growth $Ct^{1.33}$ prescribed by Eq.~\eqref{eq:ml43} regardless of the selected threshold.
In contrast, the \so solutal \os  mixing layer exhibits a strong dependence on the chosen threshold. For small thresholds, the definition captures the turbulent front, and the growth remains as  $Ct^{1.33}$. Increasing the threshold, the mixing length captures the near-interfacial dynamics, driven by the fresh, stably stratified meltwater layer.  In this case, the solutal mixing length grows diffusively as $L_\sigma \sim D_\sigma t^{1/2}$, with a prefactor $D_\sigma$ that depends on the threshold value.

These results highlight the sensitivity of mixing diagnostics to their precise  definition. Here,  different thresholds probe distinct physical processes, from turbulent transport in the bulk to diffusion-dominated dynamics near the interface. This sensitivity suggests potential caveats in more complex oceanographic settings, where mixed-layer depths are commonly defined using \sii scalar \os threshold-based criteria \citep{de2024mixed,allende2023ability}. 

\section{Concluding remarks}
\label{sec:conc}
 We have examined the  melt-driven convection beneath an ice\textendash ocean interface. Consistent with previous studies {\citep{xue2024flow}}, our setting exhibits a subdiffusive-dominated regime $\propto t^{-0.2}$ at small $R_\rho$ and a diffusion-dominated regime $\propto t^{-0.5}$ at large $R_\rho$ in the regulation of  the thermal flux at the upper boundary.
However, we find that the diffusive dynamics remain confined only to a thin interfacial layer, with  the bulk flow evolving through a (turbulent) convective thermal layer. 
Regardless of the near-interface regime, the front of the turbulent layer grows super-diffusively as $L \propto t^{4/3}$ for all $R_\rho$ and $Le$. This scaling differs from the $t^2$ growth associated with Rayleigh\textendash Taylor convection \citep{chertkov2003phenomenology,boffetta2017incompressible}, reflecting the presence of a rigid lid at the top boundary.  The melting boundary condition controls  the solutal flux in the convective mixing layer.  
\so
In particular, for large density ratios, only a fraction %input rate of salt %injected 
%in the bulk, where the fresh water input rate  
%decreasing asymptotically as 
$\propto R_\rho^{-1}$  of the meltwater input flux contributes to the solutal convective flux in the bulk.%for large density ratios, relatively to the  

While strongly idealised, our setting showcases a non-trivial interplay between a diffusive meltwater boundary  layer and a convective interior. The interfacial layer is governed by double-diffusive effects, while turbulent transport drives the  bulk.
The simplicity of the setting, which represents meltwater input as a nonlinear boundary condition without explicitly resolving phase transitions, allows to decipher the governing scaling laws. In short, the physics related to the interfacial boundary layer exhibits strong  dependence on $Le$ and $R_\rho$, while that related to convective effects is largely universal. 
Our simulations suggest a constant-flux scenario across the turbulent mixing layer, implying in particular that the subdiffusive evolution of Nusselt at the boundary $\propto t^{-0.2}$ is a finite-Reynolds effect, likely reflective of an asymptotic regime characterised by a  constant-Nusselt meltwater input.

\sii
Several questions remain relating  to  the detailed phenomenology of the flow and its dependence upon key physical parameters (dimensionality, initial stratifications, non-stationarity). 
\si  Achieving a detailed phenomenology of melting-driven turbulence, in the spirit of the work of \cite{chertkov2003phenomenology} on Rayleigh-Taylor turbulence constitutes a natural theoretical extension of the present work.
\is
Another issue relates to the flow universality, \os as the  sharp  interfacial gradient at the top interface makes the initial  configuration effectively singular, with potential implications for its statistical predictability---see, \emph{e.g.} \citep{biferale2018rayleigh,thalabard2020butterfly}.
To control the regularisation of  the initial singularity, we chose to  initiate melting through a thermal Rayleigh\textendash Taylor instability slightly offset from the interface. This physical regularisation enables  the boundary conditions to be %unambiguously and 
explicitly enforced at early times, and  triggers the convective turbulent layer. While no dependence on $\delta \to 0$, was found, it remains unclear whether such turbulent generation is universal across different regularisation schemes---in particular  implementations relying on implicit regularisations only through numerical discretisation. 
%This is a matter for future work.

%Several questions remain relating  to  the detailed phenomenology of the flow and to the universality.  \is Beyond key physical parameters (dimensionality, initial stratifications, non-stationarity), the  sharp  interfacial gradient at the top interface makes the initial  configuration effectively singular, with potential implications for its statistical predictability---see, \emph{e.g.} \citep{biferale2018rayleigh,thalabard2020butterfly}.
%To control the regularisation of  the initial singularity, we chose to  initiate melting through a thermal Rayleigh\textendash Taylor instability slightly offset from the interface. This physical regularisation enables  the boundary conditions to be %unambiguously and 
%explicitly enforced at early times, and  triggers the convective turbulent layer. While no dependence on $\delta \to 0$, was found, it remains unclear whether such turbulent generation is universal across different regularisation schemes---in particular  implementations relying on implicit regularisations only through numerical discretisation. 
%This is a matter for future work.

\section*{Acknowledgements}
SA was supported by the Marie Skłodowska-Curie Actions (MSCA) under grant agreement 101150225. This work was granted access to the HPC resources of TGCC under the GENCI allocation A0170115703 (project: Simulation Numérique des Interactions entre l'Océan et la Glace de Mer).

\bibliographystyle{jfm}
\bibliography{jfm}

@article{allende2024impact,
  title={Impact of ocean vertical-mixing parameterization on Arctic sea ice and upper-ocean properties using the {NEMO-SI3} model},
  author={Allende, S. and Treguier, A.-M. and Lique, C. and de Boyer Mont{\'e}gut, C. and Massonnet, F.  and Fichefet, T. and Barth{\'e}lemy, A.},
  journal={Geoscientific Model Development},
  volume={17},
  number={20},
  pages={7445--7466},
  year={2024},
  publisher={Copernicus Publications G{\"o}ttingen, Germany}
}

@article{allende2023ability,
  title={On the ability of {OMIP} models to simulate the ocean mixed layer depth and its seasonal cycle in the Arctic Ocean},
  author={Allende, S and Fichefet, T. and Goosse, H. and Treguier, A-M.},
  journal={Ocean Modelling},
  volume={184},
  pages={102226},
  year={2023},
  publisher={Elsevier}
}

@book{barenblatt1996scaling,
  title={Scaling, self-similarity, and intermediate asymptotics: dimensional analysis and intermediate asymptotics},
  author={Barenblatt, G.},
  number={14},
  year={1996},
  publisher={Cambridge University Press}
}

@article{berhanu2021solutal,
  title={Solutal convection instability caused by dissolution},
  author={Berhanu, M. and Philippi, J. and Courrech du Pont, S. and Derr, J.},
  journal={Physics of Fluids},
  volume={33},
  number={7},
  year={2021},
  publisher={AIP Publishing}
}

@article{boffetta2017incompressible,
  title={Incompressible {Rayleigh--Taylor} turbulence},
  author={Boffetta, G. and Mazzino, A.},
  journal={Annual Review of Fluid Mechanics},
  volume={49},
  number={1},
  pages={119--143},
  year={2017},
  publisher={Annual Reviews}
}

@article{biferale2018rayleigh,
  title={{Rayleigh-Taylor} turbulence with singular nonuniform initial conditions},
  author={Biferale, L. and Boffetta, G. and Mailybaev, A. and Scagliarini, A.},
  journal={Physical Review Fluids},
  volume={3},
  number={9},
  pages={092601},
  year={2018},
  publisher={APS}
}

@article{campolina2025non,
  title={Non-unique self-similar blowups in shell models: insights from dynamical systems and machine-learning},
  author={Campolina, C. and Simonnet, E. and Thalabard, S.},
  journal={Journal of Physics A: Mathematical and Theoretical},
  volume={58},
  number={17},
  pages={175701},
  year={2025},
  publisher={IOP Publishing}
}

@article{cohen2020buoyancy,
  title={Buoyancy-driven dissolution of inclined blocks: Erosion rate and pattern formation},
  author={Cohen, C. and Berhanu, M. and Derr, J. and Courrech du Pont, S.},
  journal={Physical Review Fluids},
  volume={5},
  number={5},
  pages={053802},
  year={2020},
  publisher={APS}
}

@article{chertkov2003phenomenology,
  title={Phenomenology of {Rayleigh-Taylor} turbulence},
  author={Chertkov, M.},
  journal={Physical Review Letters},
  volume={91},
  number={11},
  pages={115001},
  year={2003},
  publisher={APS}
}

@article{couston2024turbulent,
  title={Turbulent ice-ocean boundary layers in the well-mixed regime: insights from direct numerical simulations},
  author={Couston, L.-A.},
  journal={Journal of Physical Oceanography},
  year={2024},
  publisher={American Meteorological Society}
}

@article{de2024mixed,
  title={Mixed layer depth over the global ocean},
  author={de Boyer Mont{\'e}gut, C},
  journal={SEANOE [data set]},
  volume={10},
  pages={98226},
  year={2024}
}

@article{du2024physics,
  title={The physics of freezing and melting in the presence of flows},
  author={Du, Y. and Calzavarini, E. and Sun, C.},
  journal={Nature Reviews Physics},
  volume={6},
  number={11},
  pages={676--690},
  year={2024},
  publisher={Nature Publishing Group UK London}
}

@book{drazin2004hydrodynamic,
  title={Hydrodynamic stability},
  author={Drazin, P. and Reid, William H.},
  year={2004},
  publisher={Cambridge university press}
}

@article{favier2019rayleigh,
  title={Rayleigh--{B}{\'e}nard convection with a melting boundary},
  author={Favier, B. and Purseed, J. and Duchemin, L.},
  journal={Journal of Fluid Mechanics},
  volume={858},
  pages={437--473},
  year={2019},
  publisher={Cambridge University Press}
}

@article{gastine2025rotating,
  title={Rotating convection with a melting boundary: An application to the icy moons},
  author={Gastine, T. and Favier, B.},
  journal={Icarus},
  volume={429},
  pages={116441},
  year={2025},
  publisher={Elsevier}
}

@article{gayen2016simulation,
  title={Simulation of convection at a vertical ice face dissolving into saline water},
  author={Gayen, B. and Griffiths, R. and Kerr, R.},
  journal={Journal of Fluid Mechanics},
  volume={798},
  pages={284--298},
  year={2016},
  publisher={Cambridge University Press}
}

@article{guo2025effects,
  title={The effects of double diffusive convection on the basal melting of solid ice in seawater},
  author={Guo, R. and Yang, Y.},
  journal={Journal of Fluid Mechanics},
  volume={1013},
  pages={A24},
  year={2025},
  publisher={Cambridge University Press}
}

@article{hewitt2020subglacial,
  title={Subglacial plumes},
  author={Hewitt, I.},
  journal={Annual Review of Fluid Mechanics},
  volume={52},
  number={1},
  pages={145--169},
  year={2020},
  publisher={Annual Reviews}
}

@article{huppert1990fluid,
  title={The fluid mechanics of solidification},
  author={Huppert, H.},
  journal={Journal of Fluid Mechanics},
  volume={212},
  pages={209--240},
  year={1990},
  publisher={Cambridge University Press}
}

@article{jenkins1999impact,
  title={The impact of melting ice on ocean waters},
  author={Jenkins, A.},
  journal={Journal of physical oceanography},
  volume={29},
  number={9},
  pages={2370--2381},
  year={1999},
  publisher={American Meteorological Society}
}

@article{keitzl2016impact,
  title={Impact of thermally driven turbulence on the bottom melting of ice},
  author={Keitzl, T. and Mellado, J.-P. and Notz, D.},
  journal={Journal of Physical Oceanography},
  volume={46},
  number={4},
  pages={1171--1187},
  year={2016}
}

@article{middleton2021numerical,
  title={Numerical simulations of melt-driven double-diffusive fluxes in a turbulent boundary layer beneath an ice shelf},
  author={Middleton, L. and Vreugdenhil, C. and Holland, P. and Taylor, J.},
  journal={Journal of Physical Oceanography},
  volume={51},
  number={2},
  pages={403--418},
  year={2021}
}

@article{millero2010history,
  title={History of the equation of state of seawater},
  author={Millero, Frank J},
  journal={Oceanography},
  volume={23},
  number={3},
  pages={18--33},
  year={2010},
  publisher={JSTOR}
}

@article{peralta2015seasonal,
  title={Seasonal and interannual variability of pan-{A}rctic surface mixed layer properties from 1979 to 2012 from hydrographic data, and the dominance of stratification for multiyear mixed layer depth shoaling},
  author={Peralta-Ferriz, C. and Woodgate, R.},
  journal={Progress in Oceanography},
  volume={134},
  pages={19--53},
  year={2015},
  publisher={Elsevier}
}

@article{ramadhan2020oceananigans,
  title={Oceananigans. jl: Fast and friendly geophysical fluid dynamics on {GPUs}},
  author={Ramadhan, A. and Wagner, G. and Hill, C. and Campin, J.-M. and Churavy, V. and Besard, T. and Souza, A. and Edelman, A. and Ferrari, R. and Marshall, J.},
  journal={Journal of Open Source Software},
  volume={5},
  number={53},
  year={2020}
}

@article{rosevear2025does,
  title={How Does the Ocean Melt Antarctic Ice Shelves?},
  author={Rosevear, M.G. and Gayen, B. and Vreugdenhil, C.A. and Galton-Fenzi, B.K.},
  journal={Annual Review of Marine Science},
  volume={17},
  year={2025},
  publisher={Annual Reviews}
}

@book{talley2011descriptive,
  title={Descriptive physical oceanography: an introduction},
  author={Talley, Lynne D},
  year={2011},
  publisher={Academic press}
}

@article{vreugdenhil2019stratification,
  title={Stratification effects in the turbulent boundary layer beneath a melting ice shelf: Insights from resolved large-eddy simulations},
  author={Vreugdenhil, C. and Taylor, J.},
  journal={Journal of Physical Oceanography},
  volume={49},
  number={7},
  pages={1905--1925},
  year={2019},
  publisher={American Meteorological Society}
}

@article{xu2025aspect,
  title={Aspect ratio effect on side and basal melting in fresh water},
  author={Xu, Dehao and Yang, Rui and Verzicco, Roberto and Lohse, Detlef},
  journal={Journal of Fluid Mechanics},
  volume={1010},
  pages={A40},
  year={2025},
  publisher={Cambridge University Press}
}

@article{xue2024flow,
  title={Flow regimes in a melting system composed of binary fluid: transition from penetrative convection to diffusion},
  author={Xue, Z.-H. and Zhang, J. and Ni, M.-J.},
  journal={Journal of Fluid Mechanics},
  volume={998},
  pages={A14},
  year={2024},
  publisher={Cambridge University Press}
}

@article{yang2023ice,
  title={Ice melting in salty water: layering and non-monotonic dependence on the mean salinity},
  author={Yang, R. and Howland, C. and Liu, H.-R. and Verzicco, R. and Lohse, D.},
  journal={Journal of Fluid Mechanics},
  volume={969},
  pages={R2},
  year={2023}
}

@article{yang2025asymmetric,
  title={Asymmetric equilibrium states for melting and freezing in thermal convection},
  author={Yang, R. and Xu, D. and Verzicco, R. and Lohse, D.},
  journal={Journal of Fluid Mechanics},
  volume={1017},
  pages={A12},
  year={2025},
  publisher={Cambridge University Press}
}

@article{thalabard2020butterfly,
  title={From the butterfly effect to spontaneous stochasticity in singular shear flows},
  author={Thalabard, S. and Bec, J. and Mailybaev, A. A.},
  journal={Communications Physics},
  volume={3},
  number={1},
  pages={122},
  year={2020},
  publisher={Nature Publishing Group UK London}
}

@article{wettlaufer2001stefan,
  title={The {S}tefan problem: Polar exploration and the mathematics of moving boundaries},
  author={Wettlaufer, J.},
  journal={Die Zentralanstalt f{\"u}r Meteorologie und Geodynamik, 1851--2001, 150 Jahre Meteorologie und Geophysik in Osterreich},
  pages={420--435},
  year={2001},
  publisher={Styria Verlag Graz, Austria}
}

\appendix
\section{Details on the numerical protocol}
\label{sec:A}
\sii
This section provides additional  information regarding our numerical protocol, detailing the initial conditions and the role of $\delta$ as a smoothing numerical parameter.
\subsection{Initial conditions}
The initial salinity field is taken to be spatially uniform, $S_i(x,z) = S_\infty$. This choice prescribes both the interfacial temperature at the ice--water boundary through the freezing-point relation and the far-field temperature as, respectively,

\begin{equation}
T_{\text{top}} = T_m(S_\infty) := T_\text{f,0} + \lambda_1 S_\infty,\quad\quad T_\infty = T_{\text{top}} + \Delta T.
\end{equation}

In order to avoid an abrupt numerical adjustment at the interface, we smooth out the  temperature jump  over a thin vertical layer of thickness $\delta$. Specifically, we define $z_0 = -2\delta$, $h = \delta/4$  to construct the (unperturbed) background temperature profile as
\begin{equation}
T_i(z) =
\dfrac{
T_\infty+T_{\mathrm{top}}\exp\left(2\frac{z-z_0}{h}\right)
}{
1+\exp\left(2\frac{z-z_0}{h}\right)
},
\end{equation}
which interpolates between $T_i(-\infty) = T_\infty$ and $T_i(0) = \left(T_\infty e^{-16} + T_{top} \right)/(e^{-16}+1) \simeq T_{top}$.
This temperature field has \sii a vanishing gradient %$\simeq 0$ 
\so at the top  interface and therefore satisfies the boundary condition \eqref{eq:bc} with the constant field $S_i(x,z)=S_\infty$.
To seed instabilities, the  temperature field is perturbed  by  small-amplitude \sii Gaussian \so random noise, drawn
independently and everywhere in space as 
$$
T_i^{\epsilon}(x,z) = T_i(z) + \sqrt{\epsilon} \eta(x,z) \quad \eta(x,z) \sim \mathcal{N}\!\left(0,1\right).
$$
In our simulations, we typically set  $\epsilon = 10^{-10}$ and $\delta=0.05\,L_z$.

\sii
\subsection{Numerical implementation of the boundary conditions}
At the ice--ocean interface,  the dimensional version of the melting boundary conditions \eqref{eq:bc} involve the  top salinity $S_\text{top}=S(z=0)$ and the temperature $T_\text{top}=T(z=0)$, together with their derivatives  as
\begin{equation}
T_\text{top} = T_m(S_\text{top}) \text{~(melting point)} \quad \text{~and~}\quad \left.\frac{\partial S}{\partial z}\right|_{z=0} = Le^{-1} S_\text{top} \dfrac{St}{\Delta T} \left. \frac{\partial T}{\partial z}\right|_{z=0} \text{(flux balance)},
\label{eq:bcdim}
\end{equation}
where we recall that $Le, St$ and $\Delta T$ denote the Lewis number, the Stefan number and the temperature jump, respectively. The melting point $T_M(S) = T_\text{f,0}+\lambda_1 S$ is a function of salinity with the empirical coefficients $\lambda_1 = -5.73 \times 10^{-2}$\,K/(g/kg), and $T_{f,0} = 273.15 K$.

To enforce the boundary condition \ref{eq:bcdim} in our numerics, we discretise it with finite-difference scheme in order
to prescribe at each time step a value for $T_\text{top}$ and $S_\text{top}$ as a function of $T_{1}$ and $S_1$---representing the temperature and salinity at gridpoints \sii $z=-\Delta z$. 
 Substituting  \so the flux balance into the melting condition  yields a quadratic equation for the interfacial temperature:
\begin{equation}
	 T_\text{top}^2 + b T_\text{top} - c = 0, \quad \text{with} \quad 
	\left \lbrace
	\begin{split}
	& b(T_1,S_1) := T_1 + \frac{1}{Le}\frac{\Delta T}{St} + T_\text{f,0},  \\
& c(T_1,S_1) := \frac{1}{Le}\frac{\Delta T}{St}\lambda_1 S_1 + T_\text{f,0} \left(T_1 + \frac{1}{Le}\frac{\Delta T}{St}\right).
	\end{split}
	\right.
\end{equation}
The physically relevant root is the one prescribing the positive interfacial temperature
\begin{equation}
	T_\text{top} = \frac{-b + \sqrt{b^2 + 4c}}{2},
\end{equation}
which in turn  prescribes the interfacial salinity 
\begin{equation}
	S_{\text{top}} = \frac{T_\text{top} - T_\text{f,0}}{\lambda_1}.
\end{equation}

\section{Diffusive dynamics near the boundary}
\label{sec:B}
In our setting, the melting boundary at the  top requires vanishing normal flow. The dynamics of the temperature and the salinity therefore reduce to the diffusive dynamics
\begin{equation}
\partial _t \theta = Re^{-1} \, Pr^{-1} \Delta \theta, \quad 
\partial _t \sigma = Re^{-1} \, Pr^{-1}  \, Le^{-1}   \Delta \sigma,
\label{eq:ns-boussinesq-order}
\end{equation}
Figure~\ref{fig:thetatop} suggests that at $t\gtrapprox5$, both  the temperature and the salinity reach  interfacial values that only slowly vary in time. Treating those values as stationary \sii with respect to diﬀusive processes \so and  averaging over $x$ yields explicit solution in terms of the interfacial values:
\begin{equation}
\begin{split}
	& \left\langle \theta\right\rangle_x  = \left(\theta_0-1\right) \, \text{erfc} \left( \dfrac{ |z|}{\delta_\theta}\right), \quad 	\left\langle \sigma\right\rangle_x  = \sigma_0 \,\text{erfc} \left( \dfrac{| z|}{\delta_\sigma}\right),
\end{split}
\label{eq:BL}
\end{equation}
where $\delta_\theta(t) := Re^{-1/2} Pr^{-1/2} t^{1/2}$ and $\delta_\sigma(t) := Le^{-1/2} \delta_{\theta}$.
We use  the complementary error function  $\text{erfc}(z) = {\frac{2}{\sqrt \pi}}\int_{z}^\infty e^{-u^2/2} du$
and recall the shorthand $\sigma_0= \left\langle \sigma\right\rangle_{x,z=0}$ and $\theta_0 = \left\langle \theta\right\rangle_{x,z=0}+1$--- denoting the excess temperature.
Equation \eqref{eq:BL}  provides the gradient estimates at the interface as 
\begin{equation}
\left. \partial_z \left\langle \theta\right\rangle_x\right|_{z=0} = {\frac{2}{\sqrt\pi}}\dfrac{\theta_0-1}{\delta_\theta},\quad \left. \partial_z \left\langle \sigma\right\rangle_x\right|_{z=0} = {\frac{2}{\sqrt \pi}}\dfrac{\sigma_0}{\delta_\sigma}.
\label{eq:Derivative}
\end{equation}
Substituting those estimates into the melting condition  in Eq.~\eqref{eq:bc} yields
\begin{equation}
\theta_0 = - \gamma_1 \sigma_0,\quad \sigma_0 = -\left(\sigma_0+R_\rho\right)\,Le^{1/2}\, St \left(\gamma_1\sigma_0+1\right).
\end{equation}
For small solutal perturbations such that \sii $\sigma_0\ll R_\rho$ and $\sigma_0\ll \gamma_1^{-1}$, \so this  yields the scaling estimate
\begin{equation}
 \theta_0 =-\gamma_1 \sigma_0 \sim St Le^{1/2}R_\rho\quad;
\label{eq:scaling}
\end{equation} 
This is compatible with the scaling $O(Le^{1/2}R_\rho)$  observed in Fig.~\ref{fig:thetatop}.
%This calculation also provides the subbdominant correction  $O( LeR_\rho + \gamma_1LeR_\rho^2)$.

\si
Combined  with Eq.~\eqref{eq:Derivative}, this yields the  following (diffusive) estimate for the Nusselt number
\begin{equation}
	Nu = C_st^{-1/2},\quad C_s \simeq \dfrac{2}{\sqrt \pi}  \left( Re Pr \right)^{1/2} \left(1-Le^{1/2}StR_\rho \right).
	\label{eq:diffusive}
\end{equation}
The estimate support the dispersive nature of the diffusive constant $C_s$, with respect to 
the Lewis number $Le$ and the density ratio $R_\rho$ seen in Fig.~\ref{fig:Nu}.   
Quantitatively, it  does capture the decrease of $C_s$  with $R_\rho$ observed in panels (a,b) of Fig.~\ref{fig:Nu}. It however fails to reflect the small increase with $Le$ at high $R_\rho$, and does not yield the specific scaling exponents in the transitional region between diffusive and convective regimes. Capturing this effect would require a refined asymptotic analysis involving higher-order corrections.

\end{document}